\documentclass[11pt]{article}
\usepackage[utf8]{inputenc}
\usepackage[T1]{fontenc}
\usepackage{mathptmx}
\usepackage[scaled=0.9]{helvet}
\usepackage{courier}
\usepackage[margin=1in]{geometry}
\usepackage{amsmath,amssymb,amsthm,mathtools}
\usepackage{graphicx}
\setkeys{Gin}{keepaspectratio}
\usepackage{longtable,booktabs,array,calc}
\usepackage{float}
\usepackage{framed}
\usepackage[font=small,labelfont=bf]{caption}
\usepackage{xcolor}
\usepackage{url}
\usepackage[hidelinks]{hyperref}
\usepackage{microtype}
\providecommand{\tightlist}{\setlength{\itemsep}{0pt}\setlength{\parskip}{0pt}}

\makeatletter
\def\fps@figure{tbp}
\makeatother

\input{experiments/latex/format}
\newcommand{\FullPaper}{}
\newcounter{paperTheorem}
\newcounter{paperProposition}
\newcounter{paperListing}
\newcounter{paperAlgorithm}
\newcounter{paperCorollary}

\renewcommand{\ExperimentCitation}[1]{~\cite{#1}}

\title{Learning to Formulate: Agentic Hardware--Software Co-Design of AI Infrastructure through Progressive Optimization and Verification}
\author{Yi Ding\\Shenzhen Loop Area Institute (SLAI)\\\texttt{yiding@slai.edu.cn} \and Li Shang\\National University of Singapore\\\texttt{li.shang@nus.edu.sg}}
\date{October 27, 2026}
\hypersetup{pdftitle={Learning to Formulate: Agentic Hardware--Software Co-Design of AI Infrastructure through Progressive Optimization and Verification},pdfauthor={Yi Ding and Li Shang}}
\begin{document}
\maketitle

\begin{abstract}
AI infrastructure, the chips and software that train and serve AI models, is designed in layers. Layers make the system manageable, but they also fix which choices an optimizer can consider. Gains that need the chip and its software to change together are excluded before optimization begins: the formulation leaves them out. Language-model agents make it possible to revisit these boundaries, but an agent's proposal cannot show by itself that a design is correct or optimal. We present Formulaic, a flow that divides the work. The agent writes the formulation, a mathematical solver optimizes within it, and verification and measurement check its assumptions and results. The agent refines the problem step by step, from the computation graph toward the chip. Each step is a template: a family of programs with the conditions under which they are correct. Wall relaxation searches the chip by pricing the hardware limits that bind the software. This paper takes a first step, through separate studies rather than one end-to-end run. On NVIDIA GB10, templates the agent wrote run complete Qwen3-8B layers 1.075$\times$ faster than SGLang. In a simulated Ventus GPU study, joint design cuts the worst-scenario cycles of Qwen2.5-0.5B to 0.428 of the reference, against 0.934 for software alone. On a five-step GEMM refinement chain, reusing proofs makes checking 2.51$\times$ faster.
\end{abstract}

\hypertarget{introduction}{%
\section{Introduction}\label{introduction}}

Progress in AI is increasingly limited by the infrastructure that runs it. The cost, energy and availability of computation, memory and data movement now decide what can be trained and served. Designing that infrastructure is a co-design problem across the whole stack: the workload, the software that executes it, from serving systems through compilers and kernel libraries, and the chip underneath, down to its microarchitecture. The layers are deeply coupled. Whether a kernel runs fast depends on the register file and on-chip memory; whether that memory is worth its silicon depends on whether the software can reuse data through it. The largest gains come from exactly these interactions. Yet industry develops chips and software on separate tracks, each split layer by layer behind interfaces frozen early. A chip is architected against the workloads of the past, a tile size is chosen for a register file sized for last year's tiles, and the cross-layer combinations where specialization pays off are never explored. The mismatch keeps growing: a new model can change the demands on hardware within months, while a new chip takes years.

Layered design makes complex systems manageable, but it also fixes which interactions an optimizer can consider. Some missed opportunities therefore arise before optimization begins: they are excluded by the \emph{formulation}. Removing every boundary is not an easy fix either. Written all at once, with every choice of tiling, data movement, scheduling and hardware resource exposed together, co-design becomes a very large mixed-integer program that is hard to solve at useful scale. Layer-wise engineering is a fixed, hand-made formulation that stays solvable by giving up cross-layer interactions; the monolithic model keeps every interaction and gives up solvability. The question is how to find a formulation in between, one that exposes the cross-layer choices that matter while staying small enough to solve.

Language-model agents are already applied to parts of this problem. Agents write and optimize GPU kernels in CUDA and Triton \cite{r73,r74,r103,r104,r105}, generate and optimize RTL \cite{r106,r8} and reason about processor architectures \cite{r6}, and evolutionary coding agents have found programs and algorithms that people had not \cite{r1,r2}. Many of these systems pair the agent with strong evaluators: profilers, test suites and, in some cases, formal equivalence checks and reusable skill libraries \cite{r8}. What they usually do not produce is an explicit optimization formulation: a written statement of the choices, the constraints that keep them correct, and the cost that ranks them. Without one, it is hard to say whether a better design exists in the space the agent considered, which assumptions a design relies on, or where else an idea holds.

Language-model agents create an opportunity to revisit these boundaries, by constructing and revising formulations that expose useful cross-layer choices. Their proposals, however, cannot establish correctness or optimality by themselves. Formulaic explores a division of responsibility in which agents propose formulations, mathematical solvers optimize within them, and verification and measurement check their assumptions and consequences, sending a formulation back for repair when it fails \cite{r3,r4,r11}. We call this using the agent as a \emph{white box}. The term means that the formulation is explicit and the evidence for it can be inspected; it says nothing about the model's internals. The flow is organized around one idea. \textbf{A language-model agent refines the co-design problem step by step, from the workload down toward the chip microarchitecture. Each step is a template that a solver can optimize and a verifier can check. Because each step adds few decisions, the search and the proof can stay small.}

\textbf{Templates proposed by the agent.} A template is a design pattern packaged with the conditions under which it is correct. The same conditions bound the solver's search, give the verifier its preconditions and tell the agent where the template may be reused, and we seek the weakest such conditions \cite{r12}. Proved templates are kept, with their measurements, in an evolving population \cite{r68,r69}, so a formulation found for one problem seeds the next.

\textbf{Hardware--software co-optimization.} Progressive refinement descends from the computation graph to the microarchitecture, and each step solves a small formulation over the few decisions it adds. On the chip side, the best software presses against a few hardware limits, which we call \emph{walls}. \emph{Wall relaxation} prices each wall by re-solving with it moved one step, buys the step that recovers the most latency per unit of area, and repeats, so every chip it visits comes with the best software in the model's menu.

\textbf{Verification in the loop.} Because each step adds little, each step owes only a local proof that its finer transactions implement the coarser one. Local proofs compose back to the workload and are reused when their premises are unchanged, so verification runs inside the loop, not as an audit at the end.

\textbf{Results.} This paper takes a first step in this direction through a common template representation and complementary studies of formulation construction, hardware--software optimization and verification reuse. The studies are separate; running the whole loop on one design is future work. We first set the stage on software alone. On NVIDIA GB10 with the chip fixed, the agent's templates are on par with the industry state of the art: complete Qwen3-8B layers run 1.075$\times$ faster than SGLang (geometric mean over eight scenarios), hybrid-MoE layer groups beat SGLang in seven of eight scenarios, and against a black-box coding agent given the same caps, one run each, the flow's program is faster at every input length. Co-design then adds what software alone cannot reach. In a Ventus GPGPU study that simulates 128-token prefill and extrapolates the 512-token scenarios, letting the chip change with the software cuts the worst-scenario cycles of Qwen2.5-0.5B to 0.428 of the reference, against 0.934 for the best software-only design, a 2.2$\times$ gain under the same area ceiling. In a separate complete instruction-simulation study, a 64$\times$64 tile that is slower on the reference chip becomes part of the best design once registers, tensor shape and cache change together. Finally, verification can stay affordable inside the loop: reusing proofs speeds up checking of a five-step refinement chain by 2.51$\times$, real agent edits save less (1.25$\times$), and local checks catch the faults injected at every step.

\hypertarget{the-agentic-progressive-refinement-flow}{%
\section{The Agentic Progressive Refinement Flow}\label{the-agentic-progressive-refinement-flow}}

\hypertarget{one-iteration-of-the-flow}{%
\subsection{Overview}\label{one-iteration-of-the-flow}}

\textbf{The design problem.}\phantomsection\label{what-is-designed-and-what-it-must-compute} The input is a set \(\mathcal W\) of \emph{scenarios}, such as prefill and decode of one model, that a single chip must serve. Each scenario has a \emph{workload contract}: parameters \(p\in\mathcal P\) such as tensor shapes, allowed inputs \(U(p)\), and the acceptable behaviors \(W(p,u)\) for each input, such as outputs within the permitted numerical error. The contract is the reference. It is written independently of any optimization model, so leaving a constraint out of a formulation can never erase a requirement. A \emph{design} \(d\) picks a value for every decision from the computation graph down to the chip microarchitecture (Section~\ref{iterative-optimization-through-progressive-refinement}). Its chip part \(h\), a count and configuration for each component in a hardware library, is shared across scenarios, while its software may differ per scenario. The flow solves
\[
\min_{d}\;J(d)=\max_{w\in\mathcal W}\frac{L_w(d)}{L_w^{\mathrm{ref}}}\qquad\text{subject to}\qquad A(h)\le A_{\max},\qquad g_r(d)\le0\;\;\forall r,
\]
where \(L_w\) is the latency of scenario \(w\), \(L_w^{\mathrm{ref}}\) its latency on a reference design, \(A\) the chip area and \(g_r\) further requirements such as a power envelope or a token rate. Every term depends on both the chip and the software, so every term couples the two.

\begin{figure}[!htb]
\centering
\includegraphics[width=1\textwidth,height=\textheight]{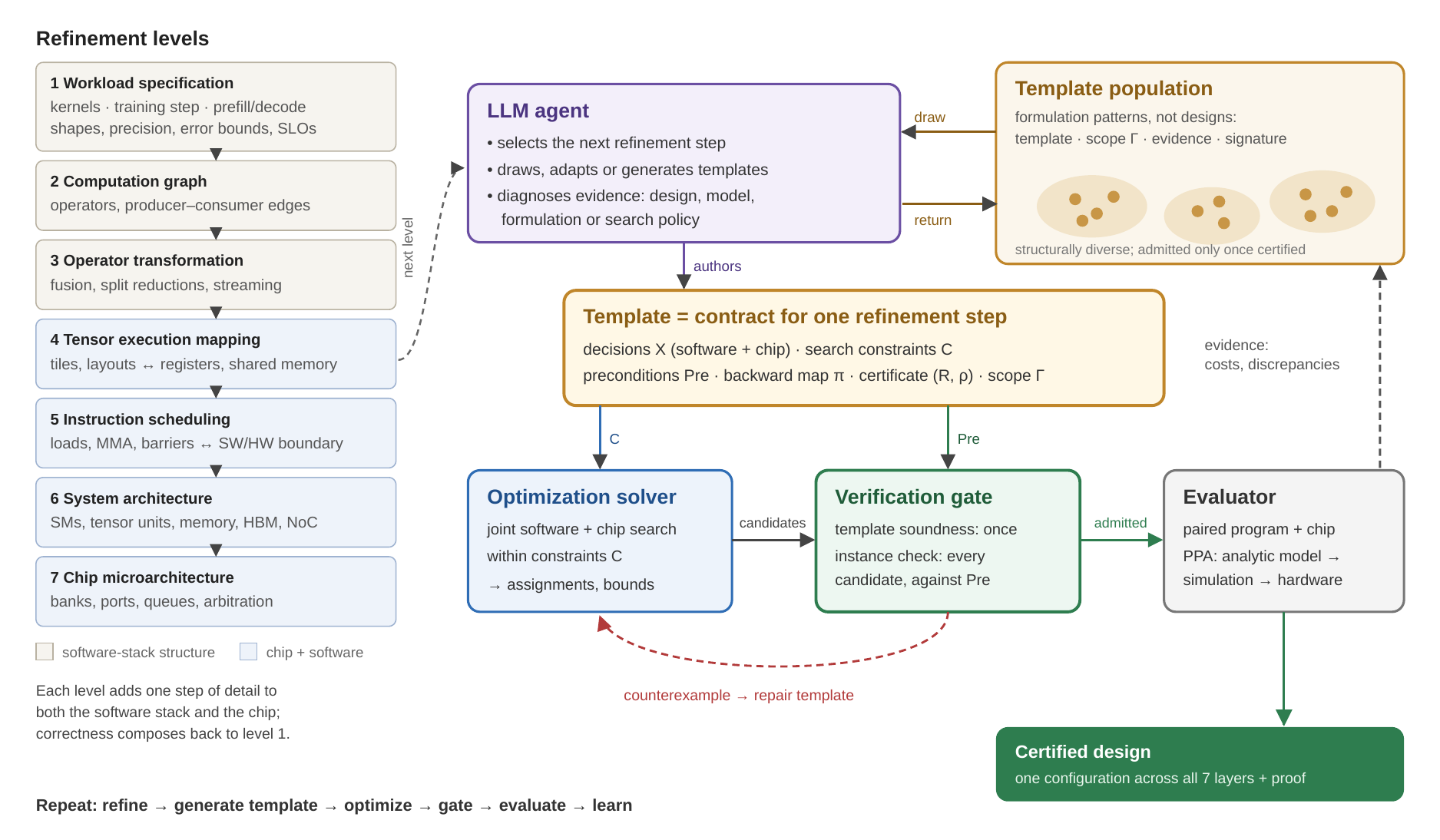}
\caption{The proposed agentic progressive refinement flow. The design moves down seven levels, from the workload specification to the chip microarchitecture, and each level adds detail to both the software and the chip. At each step the agent takes a template from the population or writes a new one. The template's search constraints \(\mathcal C\) bound the solver's joint search over software and chip decisions, and its preconditions \(\mathrm{Pre}\) are what the verifier checks for every candidate. Only candidates that pass reach the evaluator. Counterexamples go back to the agent, which repairs the template, and measured costs go back to the population. In the full flow, the output is one configuration across all levels, with a proof built from the local proof of each step. This paper tests the parts of the loop in separate studies (Section~\ref{evaluation}).}
\label{fig:flow}
\end{figure}

Solving this problem in one piece is hard, because every tile, transfer, schedule and chip resource would appear in it at once. The flow instead solves it in small steps, moving down the stack from the computation graph to the microarchitecture (Figure~\ref{fig:flow}). At each step the agent writes a template that adds only the few decisions of that step. A solver searches the template, a verifier checks it, and an evaluator measures what comes out. One iteration of this loop has five steps.

\begin{enumerate}
\def\labelenumi{\arabic{enumi}.}
\tightlist
\item
  \textbf{The agent writes a template.} It picks the next boundary to cross, for example from logical tensors to tiles in on-chip memory. It reuses a template from the population if one fits, adapts one, or writes a new one (Section~\ref{the-template-as-a-contract}).
\item
  \textbf{The verifier proves the template.} The proof is done once, for every design the template allows. If it fails, the verifier returns a counterexample and the agent repairs the template (Section~\ref{checking-contracts-the-two-tier-trust-boundary}).
\item
  \textbf{The solver optimizes.} It searches the template over software and chip decisions together, and returns the best assignment with a bound (Sections~\ref{iterative-optimization-through-progressive-refinement} and~\ref{optimization-searching-within-the-contract}).
\item
  \textbf{The choice is checked, then measured.} Because the hard proof was done in step 2, checking one assignment only means checking that it satisfies the template's constraints. The evaluator then builds the design and measures it by model, simulation or emulation.
\item
  \textbf{The agent learns.} It traces each gap between predicted and measured cost to its cause, fixes the constraints or the cost, and returns the template with its measurements to the population. Proofs whose premises did not change are kept (Section~\ref{incremental-reuse-keeping-the-gate-affordable}).
\end{enumerate}

The loop then moves one step further down the stack, or reopens an earlier decision when the measurements call for it, until the budget is spent. Its output is the best design found, the bounds that support it, and, for templates that carry proofs, a proof built from the local proofs of each step that the design computes the workload. In this paper, proofs exist for the GEMM refinement chain and the tiling template; the GB10 templates are checked numerically on tested inputs instead (Section~\ref{evaluation}). Everything in this loop turns on the template, which the next subsection defines.

\hypertarget{the-template-as-a-contract}{%
\subsection{The template as a contract}\label{the-template-as-a-contract}}

A template covers one refinement step, from level \(\ell\) to level \(\ell+1\). It introduces one new design object, and it must tell three readers what they need: the solver what it may choose and how to rank the choices, the verifier what must hold for the result to be correct, and the agent where the template can be reused. Formally it is a tuple \(\mathsf T=(O,X,\mathcal C,\mathrm{Pre},\mathcal P,\pi,\Sigma,\hat J,\Gamma)\) with five parts:
\begin{itemize}
\tightlist
\item \textbf{Scope} \(\Gamma\): the workloads, chip parameters and structural assumptions under which the template is valid, and so where the agent may reuse it.
\item \textbf{Structure} \(O,X\): the new design object, such as a shared-memory tile or an asynchronous transfer, and its decision variables, on both the software and the chip side.
\item \textbf{Search constraints} \(\mathcal C\): the feasible region over \(X\) that the solver may choose from.
\item \textbf{Correctness} \(\mathrm{Pre},\mathcal P,\pi,\Sigma\): the preconditions under which the object preserves the level-\(\ell\) computation, the operations that give it an executable meaning, a backward map to level \(\ell\), and a proof outline: a relation between fine and coarse states and a count of the work that remains (Section~\ref{backward-maps-and-local-obligations}).
\item \textbf{Cost} \(\hat J\): the cost-model terms the solver ranks designs by.
\end{itemize}

\textbf{Example: the shared-memory tiling template.} Listing~\ref{lst:tiling-contract} shows a complete template, the one evaluated in Appendix~\ref{app:contract-details}. It refines a GEMM \(C=AB\) from logical tensors to tiles staged through on-chip shared memory. Its new object is a shared-memory \emph{slot}: a thread block copies a \(T_m\times T_k\) tile of \(A\) and a \(T_k\times T_n\) tile of \(B\) into one of \(S\) slots, waits at a barrier, computes from the slot, and may wait at a second, trailing barrier (\(\beta=1\)) before the slot is refilled. Loads for reduction tile \(p+P\) are issued while tile \(p\) is being consumed, so the prefetch distance \(P\) overlaps data movement with computation. Every field of the tuple appears in the listing, tagged with the reader that uses it.

\begin{figure}[tp]
\begin{framed}
\fontsize{9}{10.5}\selectfont
\begin{verbatim}
template SharedMemoryTiling        # GEMM -> shared-memory-tiled GEMM
                                   # (operator level -> tensor-mapping level)
object O:  shared-memory slot                                          [all]
  # A block stages A[Tm x Tk] and B[Tk x Tn] into one of S slots;
  # loads of k-tile p+P overlap the computation on k-tile p.

scope Gamma:                                               [agent, verifier]
  workload   C[M,N] = A[M,K] * B[K,N],  C starts at 0,  any M, N, K >= 1
  semantics  each product A[i,k]*B[k,j] is a symbolic term; the proof shows
             every C[i,j] gets exactly its K terms (not FP32 rounding)
  chip       M_s = 96 KiB shared memory per block, 256 threads per block,
             R_acc = 32768 accumulators per block, e = 4 bytes/element
  structure  leading barrier; grid ceil(M/Tm) x ceil(N/Tn);
             tails guarded by (index < extent): loads fill 0, stores skip;
             tiles disjoint

decisions X:                                                        [solver]
  Tm, Tn in {16, 32, 64, 128, 256}          # output tile        (software)
  Tk     in {8, 16, 32, 64}                 # reduction tile     (software)
  S      in {1, 2, 3}                       # slots per block    (software)
  P      in {0, 1, 2}                       # prefetch distance  (software)
  beta   in {0, 1}                          # trailing barrier   (software)

constraints C:                                                      [solver]
  C1  S * e * (Tm*Tk + Tk*Tn) <= M_s        # shared-memory capacity
  C2  256 <= Tm*Tn <= R_acc                 # accumulator budget
  C3  P <= S - 2 + beta                     # == Pre.lifetime  (K1)

preconditions Pre:                                                [verifier]
  lifetime   P <= S - 2 + beta    # exact on the 24 checked (S, P, beta), P in 0..3

program (level l+1):                                   [verifier, evaluator]
  for each block (i, j):
    for u in 0 .. min(P, KT)-1:  load k-tile u -> slot u mod S
    for p in 0 .. KT-1:                     # KT = ceil(K/Tk)
      if p+P < KT:       load k-tile p+P -> slot (p+P) mod S
      barrier                               # leading
      compute C[i,j] += slot (p mod S)
      if beta: barrier                      # trailing

backward map pi:                                                  [verifier]
  forget the grid, the slots and the schedule  =>  C = A B over (M,N,K)

proof outline Sigma:                                              [verifier]
  relation R   slot (u mod S) holds A[:,u], B[u,:] from the completion
               of its load until the last read of k-tile u
  rank rho     loads, barriers and computes not yet executed
  to prove     coverage, exact-once, no-phantom  (index: any extent, each tile)
               readiness, lifetime  (schedule: interleavings of 2 warps x 5 phases)
               agreement: replay yields exactly the terms A[i,k]*B[k,j]

cost J^:                                                 [solver, evaluator]
  T >= T_comp,   T >= T_mem,   T >= T_comp + T_mem   if P = 0
\end{verbatim}
\end{framed}
\vspace{-0.6em}
{\small\refstepcounter{paperListing}\label{lst:tiling-contract}\textbf{Listing \thepaperListing.} The shared-memory tiling template as a contract. Tags in brackets name the party that reads each field. Constraint C3 and the lifetime precondition are the same predicate; this is condition K1. Chip parameters ($M_s$, $R_{\mathrm{acc}}$) are fixed by the scope in this instance and become decisions when a chip-level template is active. Section~\ref{checking-contracts-the-two-tier-trust-boundary} states what the proofs cover.}
\end{figure}

\textbf{Why it is a contract.} Read from top to bottom, the listing is the whole agreement between the three readers. The solver sees only the decisions \(X\), the constraints C1--C3 and the cost \(\hat J\); it never needs to know what a race is. The verifier sees only the precondition, the program, the backward map \(\pi\) and the proof outline \(\Sigma\); it never needs to know what is being optimized. The agent sees the scope \(\Gamma\): the template may be reused for any GEMM shape under the same chip parameters and structural choices, and must be looked at again when one of them changes. One rule binds the readers together. The search constraints must imply the preconditions (K1),
\[
\forall p\in\Gamma,\;\forall x\in X:\qquad \mathcal C(p,x)\;\Longrightarrow\;\mathrm{Pre}(p,x),
\]
so any design the solver picks is one the verifier has already covered. When \(\mathcal C\) is handed to a solver through auxiliary variables, the encoded constraints must imply \(\mathcal C\) as well. Two more conditions complete the contract: the preconditions must be enough to prove the step correct, once for the whole family (K2), and the constraints must allow at least one design for every parameter in scope (K3). We call a template that meets K1--K3 \emph{proved}. Checking one solution then only means confirming that it lies in \(\mathcal C\) and that its context lies in \(\Gamma\). This split, one proof for the whole family and a cheap check for each design, is what keeps verification affordable inside the loop (Section~\ref{checking-contracts-the-two-tier-trust-boundary}).

\textbf{The contract in one line.} In the tiling template the whole contract shows in one line, C3: \(P\le S-2+\beta\). It answers the question of when a slot can be refilled safely. With a trailing barrier, the slot refilled in phase \(p\) last held tile \(p+P-S\), which every warp finished before that barrier, so \(P\le S-1\) is enough. Without it, only the leading barrier of phase \(p-1\) orders a slow warp's reads, so the overwritten tile must have been used up by phase \(p-2\), which requires \(P\le S-2\). To the solver this line is a linear constraint that keeps unsafe schedules out. To the verifier it is the precondition of the lifetime condition. To the agent it is a recorded fact: the checker confirms that it is exact on all 24 checked combinations of \(S\in\{1,2,3\}\), \(P\in\{0,\dots,3\}\) and \(\beta\), in the two-warp, five-phase event model (Appendix~\ref{app:contract-details}). The check includes \(P=3\), which is outside the solver's menu, to confirm that the condition rejects it: with at most three slots, \(P\le S-2+\beta\) allows at most \(P=2\), so the menu stops there. Because C3 is the precondition itself, K1 holds by construction. K2 is the lifetime lemma together with the index conditions, proved once for the family. K3 holds because the smallest tiles with \(S=1\), \(P=0\) and \(\beta=1\) satisfy C1--C3 for every shape in scope. A listing like this is what the agent writes: a contract, not a design. We wrote this one by hand; the GB10 studies of Section~\ref{sec:exp-software} use templates the agent wrote.

\hypertarget{where-templates-come-from-exploitable-structure}{%
\textbf{How templates shrink the search.}\label{where-templates-come-from-exploitable-structure}} A contract makes a template safe to search. It does not make the search small, and it does not tell the agent which template is worth writing. Both answers come from the same place. The co-design space looks like a huge combinatorial problem, but it is not an arbitrary one. It has structure, and the structure has one source: moving data is what costs. Large memory is far away and slow to reach, so the memory hierarchy is a fixed ladder of cost, from HBM through L2 and shared memory to registers, with different bandwidth, latency and energy costs at each level. Because data movement dominates cost, in an order that is known in advance, the space has regular shapes that an expert, or an agent, can recognize and write into a template. We have found nine so far (Table~\ref{tab:structure-catalog}). They fall into three groups, according to what they do to the search.

\textbf{The first group tells the agent what to optimize.} When one resource is far costlier than the rest, it decides the cost (\emph{bottleneck dominance}). For a memory-bound projection in batch-one decode, a template can rank candidates by HBM traffic when their compute and launch costs are comparable. Performance also shows diminishing returns (\emph{concavity}). In the ideal roofline model \cite{r102}, increasing a resource helps until another resource becomes the bottleneck: more bandwidth helps a bandwidth-bound kernel, while more compute helps a compute-bound one. This directs search toward resource-balance knees and capacity boundaries, such as the largest tile that still fits in shared memory. Finally, relaxing a capacity constraint cannot worsen the model optimum when it preserves all existing choices and their costs (\emph{monotonicity}). This gives the one-sided bounds that exact pruning needs.

\textbf{The second group tells the agent what is independent.} Interactions across the stack are mostly local (\emph{near-decomposability} \cite{r101}). Shared memory sits between HBM and the registers, so the tile staged from HBM and the blocking inside the registers can be searched as two smaller problems that meet at the shared-memory tile, instead of as one three-way problem. Prefill and decode share one chip but want different software. Once the chip is fixed, each scenario is solved on its own (\emph{scenario decomposition}), and wall relaxation in Section~\ref{optimization-searching-within-the-contract} builds on this. Last, of the twenty or so decisions in a kernel, only the three or four that change how many bytes cross the bottleneck matter much (\emph{low effective dimension}). The rest are fixed or chosen coarsely.

\textbf{The third group tells the agent which points to skip.} Identical units and interchangeable memory banks produce many designs that are the same up to relabeling (\emph{symmetry}). One ordering constraint removes all copies but one. Tile sizes, slot counts and unit counts come from short menus such as powers of two (\emph{discrete menus}). The tiling template of Listing~\ref{lst:tiling-contract} has five choices for each side of the output tile and four for the reduction tile, so its domains are small enough to list. And most designs are worse than some other design on every measure (\emph{dominance}), so they can be dropped without being evaluated.

When the agent writes a template, it answers these three questions for the current refinement step: what to optimize, what is independent, and which points to skip. A template that answers none of them has no reason to exist.

\begin{longtable}[]{@{}
  >{\raggedright\arraybackslash}p{(\columnwidth - 4\tabcolsep) * \real{0.21}}
  >{\raggedright\arraybackslash}p{(\columnwidth - 4\tabcolsep) * \real{0.52}}
  >{\raggedright\arraybackslash}p{(\columnwidth - 4\tabcolsep) * \real{0.27}}@{}}
\caption{The nine structures, each with an example from a real GPU, grouped by what they do to the search.}\label{tab:structure-catalog}\\
\toprule\noalign{}
Structure & Example on a GPU & What the search can drop \\
\midrule\noalign{}
\endfirsthead
\toprule\noalign{}
Structure & Example on a GPU & What the search can drop \\
\midrule\noalign{}
\endhead
\bottomrule\noalign{}
\endlastfoot
\multicolumn{3}{@{}l}{\emph{What to optimize}} \\
Bottleneck dominance & In decode at batch one, a projection reads each weight once and does about one operation per byte read, so HBM bandwidth alone sets its time. & a full latency model: rank designs by HBM bytes \\
Concavity & A larger tile reuses more data until the kernel becomes compute-bound. Past that point it gains nothing and only uses more shared memory. & every tile size except the few near the knee \\
Monotonicity & Giving each block more shared memory never forces a slower tiling, because every tile that fit before still fits. & regions whose valid lower bounds exceed the incumbent \\
\midrule\noalign{}
\multicolumn{3}{@{}l}{\emph{What is independent}} \\
Near-decomposability & The tile staged from HBM into shared memory and the blocking inside the registers interact only through the size of the shared-memory tile. & one three-way search: two smaller ones remain \\
Scenario decomposition & Prefill of a long prompt is compute-bound and wants large tiles. Decode of one token is memory-bound and wants a split reduction. On a fixed chip each gets its own kernels. & one joint search over all scenarios \\
Low effective dimension & In the GB10 attention template, the engine, tile, split and warp choices decide the latency. Other launch details barely change it. & most knobs: fixed or coarsened \\
\midrule\noalign{}
\multicolumn{3}{@{}l}{\emph{Which points to skip}} \\
Symmetry & The SMs of a GPU are identical, so swapping which SM runs which thread block gives the same design. & copies under relabeling \\
Discrete menus & Tile sides are 16, 32, 64, 128 or 256. The tiling template of Listing~\ref{lst:tiling-contract} has only \(5\times5\times4\) tile shapes. & continuous ranges \\
Dominance & With compute and other costs equal, a tiling dominated in both HBM traffic and shared-memory demand can be omitted. & dominated designs, unevaluated \\
\end{longtable}

\textbf{From structure to template.}\phantomsection\label{two-kinds-of-obligation-correctness-gates-performance-steers} Each structure enters a template twice. For the solver it is a move in the formulation: ranking tilings by HBM bytes is a surrogate objective, splitting the search at the shared-memory tile is a decomposition, and a menu of tile sides is a short domain. For the verifier it is a condition, because most structures hold only in part of the space. Ranking by HBM bytes, for instance, is right only where the kernel is HBM-bound. Some of these conditions are cheap: symmetry, monotonicity and discrete menus are algebraic and are proved once per template. Others, such as showing that a coupling is weak, may need a small solve of their own. Unlike the preconditions of the contract, these conditions affect speed, never correctness. If one fails, the design is still correct and only the ranking was wrong, so the flow revises the cost model or reopens a branch instead of rejecting the design.

\textbf{Where the conditions point next.} A failed condition is also useful, because it says where to go next. When the evaluator reports that the bottleneck has moved, for example from HBM to shared memory, the agent writes the next template one level lower, around the data movement that now dominates. This is how the refinement steps of Section~\ref{iterative-optimization-through-progressive-refinement} arise. The capacity constraints point the other way, toward the chip. A constraint such as C1, whose right-hand side is a chip resource, is where the chip enters the software search. Section~\ref{optimization-searching-within-the-contract} calls these constraints walls and uses them to search the chip.

\hypertarget{iterative-optimization-through-progressive-refinement}{%
\subsection{Iterative optimization through progressive refinement}\label{iterative-optimization-through-progressive-refinement}}

A workload, at the top of the stack, describes its data movement only coarsely: an output \emph{is} a sum of products, and an activation \emph{flows} from one operator to the next. A running chip moves data in fine-grained transactions: a tile is copied into a numbered slot, a transfer is issued and later completes, a request waits in a queue for a bank port. Progressive refinement is the passage from the first description to the second, one step at a time. \textbf{Each refinement step replaces one coarse data transaction with finer ones that together implement it.}

Each step therefore gives the optimizer a few new decisions, such as which slot holds a tile, and gives the verifier one new condition to prove, such as that a slot is not overwritten before its last read. Neither ever faces the whole stack at once. We first lay out the levels the steps pass through, then follow one GEMM down five steps, and then show how the optimizer solves the steps in turn without losing the interactions between them.

\hypertarget{refinement-levels}{%
\subsubsection{Refinement levels}\label{refinement-levels}}

Table~\ref{tab:refinement-levels} lists the levels, from the workload specification at the top to the microarchitecture at the bottom. The workload specification is the reference; the levels below it are numbered \(\ell=0,\dots,5\), starting with the computation graph. Reading across a row gives the data transaction the level makes finer, the software and chip structure it adds, and what must be proved there. The order of the levels fixes \emph{what} becomes concrete first. It does not fix \emph{when} each choice is made, as Section~\ref{solving-a-sequence-of-formulations} explains.

\begin{longtable}[]{@{}
  >{\raggedright\arraybackslash}p{(\columnwidth - 8\tabcolsep) * \real{0.1579}}
  >{\raggedright\arraybackslash}p{(\columnwidth - 8\tabcolsep) * \real{0.2105}}
  >{\raggedright\arraybackslash}p{(\columnwidth - 8\tabcolsep) * \real{0.2105}}
  >{\raggedright\arraybackslash}p{(\columnwidth - 8\tabcolsep) * \real{0.2105}}
  >{\raggedright\arraybackslash}p{(\columnwidth - 8\tabcolsep) * \real{0.2105}}@{}}
\caption{Design objects, decisions and what must be proved at each refinement level.}\label{tab:refinement-levels}\\
\toprule\noalign{}
\begin{minipage}[b]{\linewidth}\raggedright
Level
\end{minipage} & \begin{minipage}[b]{\linewidth}\raggedright
Data transactions at this level
\end{minipage} & \begin{minipage}[b]{\linewidth}\raggedright
Software-stack structure introduced
\end{minipage} & \begin{minipage}[b]{\linewidth}\raggedright
Chip structure introduced
\end{minipage} & \begin{minipage}[b]{\linewidth}\raggedright
What must be proved
\end{minipage} \\
\midrule\noalign{}
\endfirsthead
\toprule\noalign{}
\begin{minipage}[b]{\linewidth}\raggedright
Level
\end{minipage} & \begin{minipage}[b]{\linewidth}\raggedright
Data transactions at this level
\end{minipage} & \begin{minipage}[b]{\linewidth}\raggedright
Software-stack structure introduced
\end{minipage} & \begin{minipage}[b]{\linewidth}\raggedright
Chip structure introduced
\end{minipage} & \begin{minipage}[b]{\linewidth}\raggedright
What must be proved
\end{minipage} \\
\midrule\noalign{}
\endhead
\bottomrule\noalign{}
\endlastfoot
Workload specification & Required results: what is computed, not how & Required computation, allowed inputs, operator semantics & --- & Independent reference contract \\
0. Computation graph & Tensor values flowing along graph edges & Operators, edges, parameter and state versions & --- & Matches the workload specification \\
1. Operator transformation & Reductions and partial results inside each operator & Fusion, decomposition, reduction partitioning, algorithmic families (e.g.~materialized vs.~streaming attention) & --- & Operator semantics and dependencies preserved \\
2. Tensor execution mapping & Tile copies between memory levels; values held in storage slots & Tiles, layouts, memory placement, processor mapping & Register and on-chip memory capacity, number of processing units & Coverage, exact-once reductions, address maps, value lifetime \\
3. Instruction scheduling & Issued and completed transfers; synchronized reads & Instruction selection, transfers, synchronization & Instruction interface, transfer mechanisms, software--hardware boundary & Operand readiness, visibility, progress \\
4. System architecture & Transfers across units, memory channels and interconnect & Placement of work and data across units & Execution resources, memory hierarchy, interconnect & Resource compatibility, communication contracts \\
5. Microarchitecture & Requests served by banks, ports and queues under arbitration & Bindings of operations to physical resources & Pipelines, banks, ports, queues, arbitration & Conflict handling, capacity, backpressure, justified progress \\
\end{longtable}

The chip enters at level 2, as soon as a step needs a capacity, and it enters in two ways. First, a chip resource can appear as a variable before it is fixed. Shared-memory capacity, for example, bounds the tile choice at level 2 but is chosen only at level 4, when the memory hierarchy is selected. Second, the line between software and hardware is itself a decision. A mechanism such as data transfer can be done by a software sequence, a programmable instruction or a dedicated controller. The three do the same job but differ in chip area, instruction overhead and the schedules they allow, and a mechanism built into the chip is paid for once and serves every program that uses it.

\hypertarget{a-worked-progression-one-gemm-five-steps-finer}{%
\subsubsection{One GEMM, five steps finer}\label{a-worked-progression-one-gemm-five-steps-finer}}

To see the levels at work, consider a GEMM \(Y=AB\) with dimensions \(M,N,K\), where \(Y_{ij}=\sum_{k=0}^{K-1}A_{ik}B_{kj}\). At the top, each output is a single coarse transaction: ``sum these \(K\) products.'' Five templates take it down to the hardware, and each replaces one transaction with finer ones (Table~\ref{tab:gemm-progression}). The same five steps form the refinement chain whose verification cost we measure in Section~\ref{sec:exp-verification}.

\begin{longtable}[]{@{}
  >{\raggedright\arraybackslash}p{(\columnwidth - 6\tabcolsep) * \real{0.2131}}
  >{\raggedright\arraybackslash}p{(\columnwidth - 6\tabcolsep) * \real{0.2623}}
  >{\raggedright\arraybackslash}p{(\columnwidth - 6\tabcolsep) * \real{0.2623}}
  >{\raggedright\arraybackslash}p{(\columnwidth - 6\tabcolsep) * \real{0.2623}}@{}}
\caption{A five-step GEMM refinement and the decisions exposed at each step.}\label{tab:gemm-progression}\\
\toprule\noalign{}
\begin{minipage}[b]{\linewidth}\raggedright
Step
\end{minipage} & \begin{minipage}[b]{\linewidth}\raggedright
Coarse transaction \(\to\) finer transactions
\end{minipage} & \begin{minipage}[b]{\linewidth}\raggedright
The optimizer gains
\end{minipage} & \begin{minipage}[b]{\linewidth}\raggedright
The verifier must prove
\end{minipage} \\
\midrule\noalign{}
\endfirsthead
\toprule\noalign{}
\begin{minipage}[b]{\linewidth}\raggedright
Step
\end{minipage} & \begin{minipage}[b]{\linewidth}\raggedright
Coarse transaction \(\to\) finer transactions
\end{minipage} & \begin{minipage}[b]{\linewidth}\raggedright
The optimizer gains
\end{minipage} & \begin{minipage}[b]{\linewidth}\raggedright
The verifier must prove
\end{minipage} \\
\midrule\noalign{}
\endhead
\bottomrule\noalign{}
\endlastfoot
1. Reduction partition & one sum of \(K\) products \(\to\) partial sums over groups, then a combine & the assignment \(x_{kr}\) of each reduction index \(k\) to a group \(r\) & exact-once: every product contributes exactly once \\
2. Physical storage & a logical partial result or operand \(\to\) an occupancy of a reusable buffer & buffer assignments and the order \(z\) of values sharing a buffer & lifetime: a value survives until its last read \\
3. Asynchronous transfer & an atomic write \(\to\) an issue event and a later completion & issue and completion times; the transfer mechanism (software, instruction or controller) & readiness: an operand is read only after its transfer completes \\
4. Double buffering & one buffer per value \(\to\) generations rotating through \(S\) slots, loaded \(P\) tiles ahead & slots \(S\), prefetch distance \(P\), trailing barrier \(\beta\), and the shared-memory capacity that bounds them & generation correctness: a consumer reads the right generation; lifetime becomes \(P\le S-2+\beta\) (Listing~\ref{lst:tiling-contract}) \\
5. Shared resources & an untimed transfer or access \(\to\) requests to banks, ports and bounded queues & bindings of requests to banks and ports; queue depths and port counts on the chip & conflict freedom, no deadlock, and progress under the declared arbitration contract \\
\end{longtable}

As in the tiling template, one constraint often serves both the solver and the verifier. In step 2, with event times in a horizon \([0,H]\), a declared separation \(\tau>0\) and an order variable \(z\in\{0,1\}\) for two values sharing storage,
\[
l_v+\tau\le t_{v'}+(H+\tau)(1-z),\qquad l_{v'}+\tau\le t_v+(H+\tau)\,z,
\]
where \(t\) is a value's first destructive write and \(l\) its last required read. These inequalities are the solver's constraints and the verifier's lifetime precondition at once. Step 4 applies the same idea to rotating slots, which gives the line \(P\le S-2+\beta\) of Listing~\ref{lst:tiling-contract}.

Read down the table, the optimization variables grow from a partition to a storage assignment, an event schedule, a slot rotation and a resource configuration, and the chip enters as soon as a step needs a capacity, a mechanism or a port. Read across it, every new freedom arrives with exactly the condition it could break, and none of these conditions can even be stated one step earlier. A lifetime is meaningless before there is storage, and a deadlock is meaningless before there are queues. Because each step proves only its own condition, the five proofs chain back to the single GEMM at the top (Theorem~\ref{thm:composition}). Section~\ref{verification-certifying-each-refinement} gives the proofs. The rest of this subsection turns to the optimizer.

\hypertarget{solving-a-sequence-of-formulations}{%
\subsubsection{Solving one step at a time}\label{solving-a-sequence-of-formulations}}

The optimizer moves down the stack in step with the verifier. It never writes down the monolithic program of Section~\ref{introduction}, in which every tiling, transfer, schedule and hardware resource appears at once. Instead it solves a \emph{sequence} of formulations, each adding only the decisions of one step. For the GEMM, the first formulation chooses a reduction partition. The next adds buffer assignments, then transfer times and a transfer mechanism, then a slot rotation, and last the bank, port and queue bindings. Each formulation is meant to be small. We show this for the formulations we report, not in general: the tiling step of Listing~\ref{lst:tiling-contract}, for instance, is a MILP with 93 variables that HiGHS solves to proven optimality in under 0.2~s (Appendix~\ref{app:contract-details}). Progressive refinement organizes the search; it does not by itself make every step tractable.

Solving one step at a time would be a greedy, layer-by-layer pass if every earlier decision were final. It is not. When a step begins, each earlier decision is treated in one of three ways. It is \emph{fixed} when nothing at the new step depends on it. It is \emph{carried forward as a variable} when it is coupled to the new decisions, as a tile shape is coupled to the shared-memory capacity that bounds it; this is how a chip resource can stay open across several steps. And it is \emph{reopened} when measurements show that a fixed decision is now costing performance.

\phantomsection\label{sources-of-design-loss}Measurements also say what kind of repair is needed, because a chosen design can fall short of the best one for four different reasons (Table~\ref{tab:design-loss}). The template may be unable to express the best design. Its correctness conditions may be stronger than needed, so they exclude safe designs (the precondition loss \(\Delta_{\mathrm{pre}}\) of Section~\ref{the-weakest-contract}). The solver may stop short of the model's optimum, a gap \(\delta_k\). Or the cost model may predict wrongly, an error \(\varepsilon_k\). A zero solver gap rules out only the third. It says nothing about whether the template can express a good design or whether the model predicts the chip. The Ventus model-fidelity study illustrates this distinction: zero-gap solutions still miss the simulated costs (Appendix~\ref{app:model-fidelity}). If the cost model is accurate to within \(\varepsilon_k\) on every design in the search space, the four parts bound the loss of the chosen design \(\hat d_k\): \(J(\hat d_k)-J^\star\le L_{\mathrm{rep}}+\Delta_{\mathrm{pre}}+\delta_k+2\varepsilon_k\), where \(L_{\mathrm{rep}}\) is the gap between the best design the template can express and the best design overall. Errors measured on sampled designs do not give such a uniform accuracy, so we use the four parts to sort causes, not as a guarantee.

\begin{table}[!htb]\centering\small
\caption{Four sources of design loss and the repair each calls for.}\label{tab:design-loss}
\begin{tabularx}{\linewidth}{@{}l>{\raggedright\arraybackslash}X>{\raggedright\arraybackslash}X@{}}\toprule
Source & What goes wrong & Repair\\\midrule
Representation & The template cannot express the best design & A new or revised template\\
Correctness constraints & The preconditions exclude safe designs & A less restrictive condition that is still sound\\
Optimization & The solver stops short of the model optimum & A better encoding or search\\
Prediction & The cost model ranks designs wrongly & A better model, fitted to measurements\\\bottomrule
\end{tabularx}
\end{table}

\phantomsection\label{the-agents-actions-and-how-it-chooses-them}Reopening pays off only if the right decisions are released together. Many decisions follow from a few structural ones, and the useful groups are those tied by one mechanism. An example comes from the complete instruction-simulation study on Ventus in Section~\ref{sec:exp-codesign}. There the chip was first fixed at that study's reference design, and the software was tuned for it. A 64-by-64 tile was then added to the software menu. It needs 173 vector registers, so on the reference chip it is slower than the smaller tiles, and tuning the software alone gives no reason to pick it. Enlarging the register file alone looks wasteful too, because the current tiles cannot use the extra registers. When register capacity, tensor shape and cache allocation were opened together with the tile, the large tile became part of the best design: at 1.081~mm\(^2\), the balanced design in that study reaches a 2.93\(\times\) worst-scenario speedup over its reference. These numbers come from that study's own menus and reference chip, so they should not be mixed with those of Table~\ref{tab:codesign}. The study opened these choices by design; we do not claim the agent found the combination on its own. The lesson is that the agent should reopen coupled groups, not single variables, and Section~\ref{optimization-searching-within-the-contract} turns the same idea into a search over the chip.

\hypertarget{the-evolving-template-population}{%
\subsection{The evolving template population}\label{the-evolving-template-population}}

A template that took effort to find should not be found again. The flow therefore keeps every proved template in a \emph{population}, and the agent's first move at each step (step 1 of Section~\ref{one-iteration-of-the-flow}) is to look there. Each template is stored with what the agent needs to judge whether it applies: its scope, its proofs, the designs it produced and their measurements, the counterexamples it met, and a short description of the kind of design it produces, such as its algorithm, its data reuse, where it places data, and the chip resources it relies on. At each step the agent draws the templates whose scope covers the current problem, adapts or combines them, and returns those that pass verification, together with what the iteration taught.

Three rules keep the population trustworthy as it grows. First, a template joins only after it is proved (K1--K3). Unproved templates may be tried, but their results are never used in a correctness claim. Second, every performance fact is stored with the chip and workload on which it was measured. ``Fusing these two operators is faster'' is a fact about one chip, so a later change to the chip reopens it. Third, the population keeps templates that differ in structure rather than many variants of one. An option that looks poor on today's chip, like the 64-by-64 tile on the reference chip, is then still there when the chip changes.

Over many problems, the population records which structures of Table~\ref{tab:structure-catalog} hold for which workloads on which chips. This is how the agent gets better at formulating: it starts each new problem from templates that were proved and measured before, not from scratch. In this paper the GB10 studies test templates the agent wrote and then reused on new inputs and models (Sections~\ref{sec:exp-software} and~\ref{sec:exp-efficiency}); long-term growth of the population is future work.

To sum up: for a fixed chip, the flow gives software that is optimal within its templates and cost model, and proved where the template carries a proof. The next section uses that solve to search the chip itself.

\hypertarget{optimization-searching-within-the-contract}{%
\section{Hardware--Software Co-Optimization by Wall Relaxation}\label{optimization-searching-within-the-contract}}

Section~\ref{the-agentic-progressive-refinement-flow} finds the best proved software for a given chip. This section lets the chip change too. Today the two are designed in silos: architects fix the chip from the workloads of the past, and software is then tuned on whatever chip results. Each side treats the other as fixed, so a choice that only pays when both sides change together is never found. We instead write co-design as one problem. Let \(s\) be the software decisions and \(h\) the chip configuration, with latency \(L(s,h)\), area \(A(h)\) and budget \(A_{\max}\), and let \(\mathcal S(h)\) be the software that is legal on \(h\): tiles that fit shared memory, register demand that does not spill, thread counts the scheduler accepts. Co-design is
\[
\min_{h:\,A(h)\le A_{\max}}\;\min_{s\in\mathcal S(h)}\;L(s,h).
\]
Written this way, everything that couples the two sides is visible. Section~\ref{wall-relaxation} shows that the coupling has a simple form, which we call walls, and Section~\ref{wall-algorithm} uses it to solve the whole problem with one algorithm, wall relaxation.

\hypertarget{wall-relaxation}{%
\subsection{One problem, coupled through walls}\label{wall-relaxation}}

When the chip menu is small, this problem can be solved directly as one mixed-integer program over both sides, and Section~\ref{sec:exp-codesign} does so on Ventus. That does not scale to a configurable hardware library, and when it finishes it reports a chip without saying why that chip was chosen. We therefore split the problem at the point where the two sides meet.

\paragraph{Inner and outer problems.} The \emph{inner problem} is the software optimum for a fixed chip,
\[
L^\star(h) \;=\; \min_{s\in\mathcal S(h)} L(s,h).
\]
This is exactly what the flow of Section~\ref{the-agentic-progressive-refinement-flow} already solves: the template's constraints define $\mathcal S(h)$, the solver returns $s^\star(h)$ at zero gap, and the verifier checks it. For the tiling template of Listing~\ref{lst:tiling-contract}, one such solve takes under a second (Appendix~\ref{app:contract-details}). The \emph{outer problem} chooses the chip, and it sees the software only through $L^\star(h)$. To search the chip well, we need to know what $L^\star$ looks like, and the answer lies in how a chip constrains its software.

\paragraph{Walls.} Every template in this paper bounds the software by inequalities whose right-hand sides are chip resources: shared-memory capacity $S\,e\,(T_mT_k+T_kT_n)\le M_s$, an accumulator budget $T_mT_n\le R_{\mathrm{acc}}$, and so on. The accumulator budget is per block; with $n_t$ threads and $a_{\max}$ accumulators per thread, $R_{\mathrm{acc}}=n_ta_{\max}$. We call each such inequality $g_i(s)\le h_i$ a \emph{wall}: a hardware resource seen from the software side. Walls shape the search in three ways. Most of the raw software space lies behind them; on GB10, the capacity constraints alone remove 61--64\% of the 1,296-point tiling grid. Crossing a wall is a cliff, not a slope: one step past the register wall spills, and one step past the shared-memory wall does not launch. And the best designs tend to sit at a wall, reflecting the diminishing returns and capacity limits discussed in Section~\ref{where-templates-come-from-exploitable-structure}: the largest tile that still fits, the deepest pipeline that still leaves registers. The consequence is the central point of this section. \emph{Hardware acts on software mainly by moving walls.} For capacity-only changes at fixed occupancy, enlarging the register file leaves an admitted kernel's cost unchanged but can make a larger tile legal. Walls are therefore the shared language of the two sides: the software sees them as constraints, and the chip sees them as resources to buy.

\paragraph{Shadow prices.} If hardware acts by moving walls, the natural question for the outer problem is what each wall is worth. Let $h+e_i$ be the chip identical to $h$ except that resource $i$ takes the next value in its menu, at area cost $\Delta A_i(h)=A(h+e_i)-A(h)$. The \emph{shadow price} of wall $i$ at $h$ is the latency the software recovers when the wall moves one step,
\[
\lambda_i(h) \;=\; L^\star(h) - L^\star(h+e_i), \qquad \eta_i(h) \;=\; \lambda_i(h)/\Delta A_i(h),
\]
and $\eta_i$ is its price in cycles per unit area. The name follows linear programming, where the shadow price of a constraint is read from the optimal dual. The inner problem here is integer and has no such dual, so the price is computed by re-solving the inner problem on the changed chip.

Pricing every wall this way would be wasteful, so the flow first screens out walls that cannot help. In linear programming a constraint with slack has a zero price, but with discrete menus that is not quite true: a wall with slack at $s^\star(h)$ can still exclude a better design that lies far beyond it. The safe screen is to remove the wall entirely. Let $L^\star_{-i}(h)$ be the inner optimum with wall $i$ removed.

\refstepcounter{paperProposition}\label{prop:wall-pricing}\textbf{Proposition \thepaperProposition (wall pricing).} For capacity-only walls with nested feasible sets and a hardware-independent cost $L(s)$ for each admitted program, for every wall $i$ and relaxation $t\ge 0$,
\[
0 \;\le\; L^\star(h)-L^\star(h+t\,e_i) \;\le\; \Lambda_i(h) \;=\; L^\star(h)-L^\star_{-i}(h).
\]
In particular, if $\Lambda_i(h)=0$, no relaxation of wall $i$ alone improves the design.

\emph{Proof.} $\mathcal S(h)\subseteq\mathcal S(h+t\,e_i)\subseteq\mathcal S_{-i}(h)$, where $\mathcal S_{-i}$ drops wall $i$; the minimum of $L$ is non-increasing along this chain. \(\square\)

Bottleneck dominance motivates concentrating hardware search on a few resource limits. The screen costs one inner solve per wall and tests whether removing that wall improves the software optimum. We write $B(h)$ for the walls that pass and price only these walls.

\paragraph{Staircases and joint steps.} A price is not a smooth quantity. Because the software menus are discrete, $L^\star(h+t\,e_i)$ is a non-increasing \emph{staircase} in $t$: flat until the relaxation allows a new menu point, then a drop. A price therefore belongs to a step, not a slope, and a gradient of a continuous relaxation is wrong in a specific way: zero on every flat segment and unbounded at every step. Two consequences shape the algorithm. A wall can have a zero first step and a large second one, so each binding wall is priced $k$ menu points out ($k=2$ by default). And steps of different walls interact: the tile that one more shared-memory bank unlocks may also need more accumulator registers, so a joint relaxation can be worth more than the sum of its parts. The 64-by-64 tile of Section~\ref{the-agents-actions-and-how-it-chooses-them} is such a joint step seen from the hardware side: it helps only when register capacity, tensor shape and cache allocation move together. It motivates coupling; it is not evidence that the procedure below finds such steps.

\paragraph{What the screen can miss.} Proposition~\ref{prop:wall-pricing} assumes capacity walls with nested feasible sets and a program cost that does not depend on the hardware. Walls that change throughput are outside it. Even within it, the screen tests one wall at a time, so it can miss a relaxation that only pays when two walls move. Take two walls, each at capacity 1. Program A needs \((1,1)\) and costs 10; program B needs \((2,2)\) and costs 1. Removing wall 1 alone still leaves B blocked by wall 2, and the same holds the other way, so \(\Lambda_1=\Lambda_2=0\) and neither wall passes the screen, although moving both makes the design ten times faster. Algorithm~\ref{alg:wall-relaxation} prices pairs only among walls that pass the screen, so it misses this case. The \emph{coupled} variant evaluated in Section~\ref{sec:exp-codesign} differs from Algorithm~\ref{alg:wall-relaxation} in one point: it also screens each pair of walls with both removed, and prices the pairs that pass.

\hypertarget{wall-algorithm}{%
\subsection{The algorithm and its guarantees}\label{wall-algorithm}}

\begin{figure}[t]
\small
\hrule\smallskip
\refstepcounter{paperAlgorithm}\label{alg:wall-relaxation}\textbf{Algorithm \thepaperAlgorithm: Wall relaxation.}\par\smallskip\hrule\smallskip
\textbf{Input:} scenarios $W$; hardware library with menus and area model $A$; budget $A_{\max}$; start $h_0$ with $A(h_0)\le A_{\max}$; inner solve $\textsc{Solve}(h)\mapsto(s^\star(h), L^\star(h), B(h))$, where $B(h)$ is the set of walls with $\Lambda_i(h)>0$.\par
\begin{enumerate}\setlength\itemsep{1pt}
\item $h\leftarrow h_0$; call $\textsc{Solve}(h)$; record $(h,s^\star,L^\star,A(h))$.
\item \textbf{Price.} Screen every wall by $\Lambda_i(h)$. For each $i\in B(h)$ and $t=1,\dots,k$, call $\textsc{Solve}(h+t\,e_i)$ and record the staircase. For each pair $i,j\in B(h)$, call $\textsc{Solve}(h+e_i+e_j)$ once.
\item \textbf{Buy.} Each priced move, a single step, a $t$-step move or a pair, leads to a chip $h'$. Its gain per area is $g(h')=\bigl(L^\star(h)-L^\star(h')\bigr)/\bigl(A(h')-A(h)\bigr)$. Among moves with $A(h')\le A_{\max}$, take the one with the largest $g$; break near-ties toward the smaller area. If no move has a positive gain, stop.
\item \textbf{Validate.} Measure the purchased design on the evaluator. Compare the measured latency drop with the predicted drop $L^\star(h)-L^\star(h')$. If they differ by more than the noise band, treat it as a cost-model error: recalibrate the responsible term, re-solve the incumbent $h$ under the new model, discard all stored prices, and return to step~2.
\item \textbf{Reclaim.} For each wall with slack at $h'$, try one step down, $h'-e_j$; keep it if $L^\star$ is unchanged.
\item $h\leftarrow h'$; repeat from step~2 until step~3 stops or the iteration budget $T$ is spent.
\end{enumerate}
\textbf{Output:} the visited designs $(h,s^\star(h),L^\star(h),A(h))$, their non-dominated subset, and the ordered list of purchases with the mechanism behind each step.\par\smallskip\hrule
\end{figure}

\paragraph{The algorithm.} Algorithm~\ref{alg:wall-relaxation} puts these pieces together. It starts from a chip within budget and solves its software. It then screens and prices the binding walls, $k$ steps out and in pairs, and buys the move with the largest latency drop per unit of added area. It measures the purchased design, so that a wrong price is caught as a cost-model error rather than trusted. It gives back any resource whose wall has slack, and repeats until no purchase helps or the budget is spent. The inner solve supplies both the optimum and every price. With several scenarios, the inner problem splits by scenario at fixed $h$, and a price is the drop in the worst-scenario objective of Section~\ref{what-is-designed-and-what-it-must-compute}; a wall that pays for decode but not for prefill shows up as such.

The loop itself is mechanical, and that is deliberate. The agent works at its edges. It writes templates whose walls take the chip as parameters rather than fixed values, which is what makes $\textsc{Solve}(h)$ possible. It reads each staircase and records the mechanism behind each step, such as the tile that became legal, as the reason a wall was worth buying. And when a measurement disagrees with every term of the model, it concludes that a wall is missing and adds it to the template.

\paragraph{Guarantees.} Every chip the loop visits comes with its inner optimum, optimal within the cost model and the software menu; in the full flow, that software is also checked by the verifier. The output is therefore a set of chips, each paired with the best software the model knows for it. The outer rule, however, is greedy, and Proposition~\ref{prop:wall-greedy} states what its accepted steps guarantee.

\refstepcounter{paperProposition}\label{prop:wall-greedy}\textbf{Proposition \thepaperProposition (outer loop).} For a fixed cost model, accepted positive-price purchases and objective-preserving reclaim steps produce a non-increasing sequence $L^\star(h_0),L^\star(h_1),\dots$. At each visited chip, the algorithm evaluates the specified single-wall steps and pairs of walls in $B(h)$.

The first claim follows directly from the acceptance rules; the second from the pricing step. This establishes improvement along the visited path, not a global approximation bound. Budgeted-submodular guarantees require the corresponding assumptions and additional selection or enumeration procedures \cite{khuller1999,sviridenko2004}; they do not follow for Algorithm~\ref{alg:wall-relaxation} from submodularity alone. Its unexplored choices include longer steps, pairs excluded by screening, higher-order combinations and alternative purchase paths. Recalibration changes the objective model. After it, the incumbent and every price are recomputed under the new model, so the non-increasing sequence holds within each stretch of fixed model, not across recalibrations. There is no global optimality guarantee.

\paragraph{Cost.} With $n$ walls, at most $b$ passing the screen, look-ahead $k$ and $T$ pricing rounds, the loop uses at most $1+T\bigl(2n+bk+\binom{b}{2}\bigr)$ inner solves, counting an initial solve, screening, single and paired pricing, and at most $n$ reclaim trials per round. For $n=6$, $b=4$, $k=2$ and $T=20$, this bound is 521 solves. Each purchase requires an evaluator run; recalibration can add further work. Purchase selection requires useful estimates of relative gains per area, and step~4 checks the predicted gain against measurement.

Each chip that wall relaxation visits thus comes with software that is optimal for it within the model. The next section shows how that software, and every step that produced it, is proved to compute the workload.

\hypertarget{verification-certifying-each-refinement}{%
\section{Verification in the Loop}\label{verification-certifying-each-refinement}}

In the full flow, Sections~\ref{the-agentic-progressive-refinement-flow} and~\ref{optimization-searching-within-the-contract} rely on the verifier at three points. Every template is proved before the solver may use it, every candidate the solver proposes is checked before it is kept, and every chip that wall relaxation visits is paired with software proved correct for it. The verifier therefore works inside the optimization loop, not as an audit after it, and it must be fast enough to keep up.

What makes this possible is that verification moves down the stack with the refinement, one step at a time. At every step the verifier answers one local question: do the finer transactions introduced here implement the coarser transaction they replace? It may assume that the level above has already been proved, and it owes nothing about the levels below, which have not been written yet. For the GEMM of Section~\ref{a-worked-progression-one-gemm-five-steps-finer} this means five small proofs, one per step, and none of them mentions the whole stack. In this paper we build these proofs for the five-step GEMM chain and the tiling template; the GB10 programs of Section~\ref{sec:exp-software} are checked numerically against a reference on the tested inputs, not proved. We first show how one step is proved and how the proofs compose (Section~\ref{backward-maps-and-local-obligations}), then how a template is proved once so that each candidate is cheap to check (Section~\ref{checking-contracts-the-two-tier-trust-boundary}), why the template's preconditions should be as weak as possible (Section~\ref{the-weakest-contract}), and how proofs are reused as the design changes (Section~\ref{incremental-reuse-keeping-the-gate-affordable}).

\hypertarget{backward-maps-and-local-obligations}{%
\subsection{One step, one local proof}\label{backward-maps-and-local-obligations}}

\phantomsection\label{formulations-as-families-of-executable-designs}\phantomsection\label{soundness-of-one-refinement}A refinement step relates a finer design to the coarser one it came from. The template's backward map \(\pi\) reads a level-\((\ell+1)\) design as a level-\(\ell\) design, for example recovering logical tensors from tiles and buffers, and must map every feasible fine design to a feasible coarse one. The claim to prove is that every behavior of the fine design is a behavior of the coarse one. The template's proof outline supplies a relation \(R\) between fine and coarse states and a rank \(\rho\) with values in the natural numbers that counts remaining work, and the proof splits into four local conditions:
\begin{itemize}
\tightlist
\item \textbf{Initialization:} every fine start state is related to a coarse start state.
\item \textbf{Step matching:} from related states, every fine step is explained by zero or more coarse steps with the same visible events, and the new states are again related; a step that makes no coarse progress lowers the rank.
\item \textbf{Progress:} a related fine state that has not finished can always take a step.
\item \textbf{Completion agreement:} when the fine design finishes, the coarse one has finished with the same result.
\end{itemize}
The claim covers \emph{every} design the formulation allows, for every parameter and input in scope, not only the one the solver picks. Progress rules out deadlock: a stuck state that has not finished. It does not by itself ensure that every waiting request is eventually served. That needs an explicit assumption about arbitration, for example that an arbiter grants every request within a bounded number of cycles, and step 5 of the GEMM chain declares one.

\textbf{The conditions along the GEMM chain.} The four conditions are the same at every step. What changes from step to step is the relation \(R\), and which of the four conditions is hard. In the reduction partition (step 1), \(R\) relates the partial sums to the full sum, and completion agreement is the hard part: the combined partial sums must contain every product exactly once. In physical storage (step 2), \(R\) states that every buffer holds the logical value it currently stands for, and step matching for a read is exactly the lifetime condition: the value read must be the one the coarse step consumes. In the asynchronous transfer (step 3), the coarse write is matched to the transfer's completion and the issue is internal work, so step matching fails precisely when a read comes before completion; this is the readiness condition. In double buffering (step 4), \(R\) tracks which generation each slot holds, and step matching requires that a consumer read the generation the coarse program names. In shared resources (step 5), step matching holds easily and the weight falls on progress and service: progress excludes deadlock, and the declared arbitration assumption ensures that a waiting request is eventually served.

Each step thus proves one small thing. The following theorem says that these small proofs add up to the whole.

\phantomsection\label{composition-across-the-stack}\refstepcounter{paperTheorem}\label{thm:composition}\label{thm:local-refinement}\textbf{Theorem \thepaperTheorem (composition).} Suppose that every refinement step satisfies the four conditions with a rank in the natural numbers and maps feasible designs to feasible designs; that every maximal coarse execution terminates; that the environment assumptions of adjacent levels are compatible, so that what one level assumes about its environment, such as arbitration, is guaranteed by the level below or declared as an assumption of the whole result; and that the computation graph meets the workload contract. Then every design allowed at the bottom of the stack meets the workload contract.

\emph{Proof sketch.} Each step's conditions give behavior inclusion between adjacent levels: step matching maps a fine run to coarse runs with the same visible events while keeping the states related; because the rank is a natural number, a fine run can do only finitely much work without coarse progress, so with terminating coarse runs every fine run terminates; progress rules out a fine run that stops early; and completion agreement matches the results. Chaining the inclusions down the stack gives the claim. \(\square\)

This is what lets the flow verify one step at a time. Adding a step costs one new local proof, not a new end-to-end argument. For the GEMM, the five local proofs compose into a proof that the design described in terms of banks, ports and queues computes \(Y=AB\). Correctness composes; optimality does not, unless a dominance argument is supplied.

\hypertarget{checking-contracts-the-two-tier-trust-boundary}{%
\subsection{Prove the template once, check each candidate cheaply}\label{checking-contracts-the-two-tier-trust-boundary}}

Proving each step for every design the template allows is the expensive part. The flow does it once per template, so that each candidate the solver proposes needs only a cheap check.

\textbf{What a proof covers.} Every proof has a stated domain, and for the tiling template it is narrower than the listing's workload line suggests. The index conditions (coverage, exact-once, no phantom) are proved for every tensor extent, one query per menu tile size. The schedule conditions (readiness, lifetime) are proved for every interleaving of a bounded event model with two warps and five reduction phases; larger blocks are not covered. Arithmetic is symbolic: the checker treats each product \(A_{ik}B_{kj}\) as a token and shows that each output receives exactly its \(K\) tokens. This is exact term coverage over real numbers. It does not show that two FP32 summation orders give the same result, which the GPU studies check numerically instead. The trusted base is the Python condition builder, the independent replay interpreter and the Z3 solver.

\textbf{Once per template.} When the agent writes or revises a template, the checker proves its conditions over the template's whole family, using an SMT solver for arithmetic and array conditions and library lemmas for symbolic identities. The result is one of three. \emph{Proved}: the result is stored with its premises for reuse. \emph{Counterexample}: an independent interpreter replays it, and the agent repairs the template. \emph{Unresolved}: the template stays unproved, because an open question is never treated as a proof. The agent may propose maps, relations and proof steps, but it cannot turn an unproved statement into an assumption.

\textbf{Once per candidate.} Because the template's search constraints imply its preconditions (K1), checking a solver's choice reduces to evaluating its constraints on that choice, plus a check that the solver's encoding implies the declared constraints. This is the check inside the loop, and it costs milliseconds: about 25~ms for the tiling template, against a one-time proof over the whole family.

\hypertarget{the-weakest-contract}{%
\subsection{The weakest contract}\label{the-weakest-contract}}

\begin{figure}
\centering
{\setlength{\unitlength}{\dimexpr\textwidth/900\relax}%
\begin{picture}(900,390)
\put(0,0){\includegraphics[width=\textwidth]{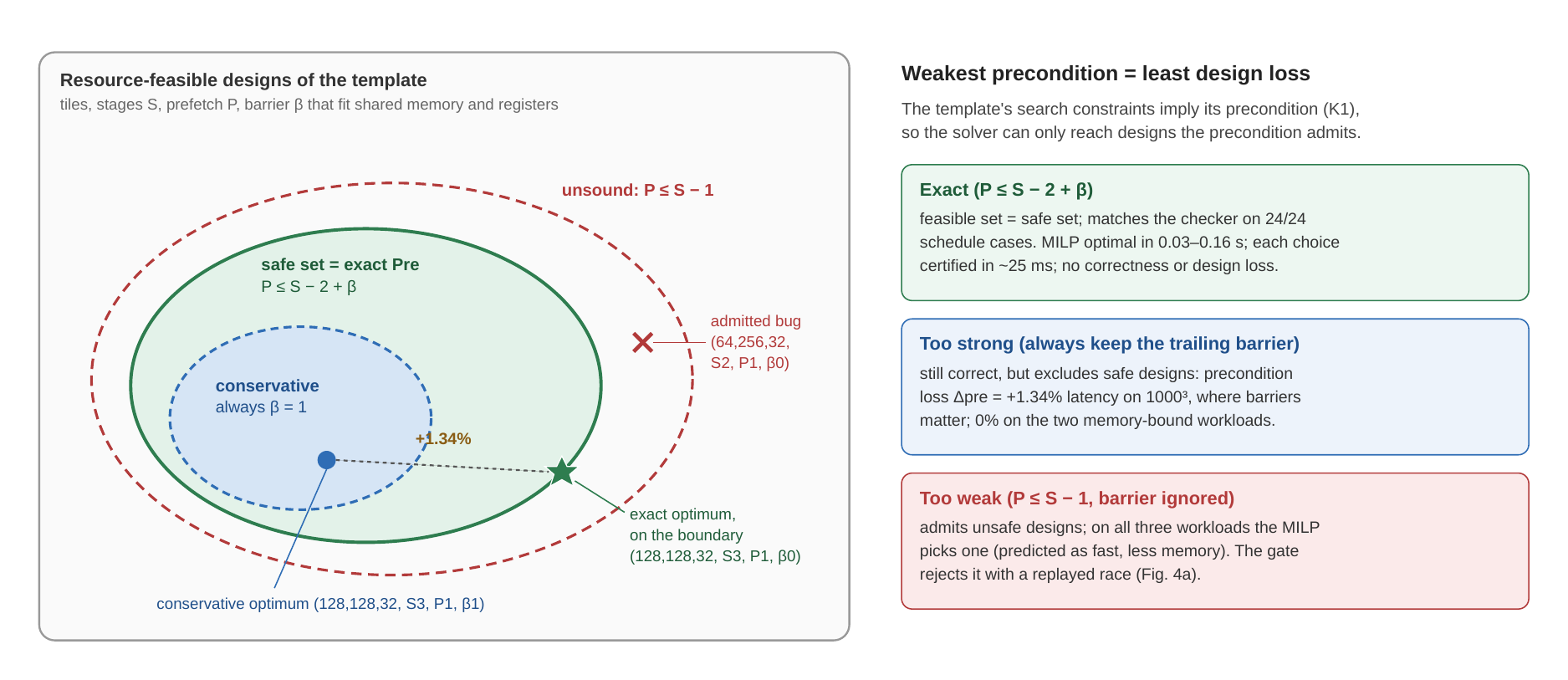}}
\definecolor{weakbg}{RGB}{251,234,234}\definecolor{figtext}{RGB}{51,51,51}%
\put(706,67){\textcolor{weakbg}{\rule{44\unitlength}{13\unitlength}}}
\put(708.6,70.6){\textcolor{figtext}{\fontsize{5.3}{6}\sffamily\selectfont The check}}
\put(650,53){\textcolor{weakbg}{\rule{46\unitlength}{14\unitlength}}}
\put(652,57.1){\textcolor{figtext}{\fontsize{5.3}{6}\sffamily\selectfont (Fig.~\ref{fig:race}a).}}
\end{picture}}
\caption{Weakest precondition as least design loss, for the shared-memory tiling template. The exact precondition equals the safe set and contains the optimum on its boundary. A conservative precondition (always keep the trailing barrier) is sound but excludes that optimum, a precondition loss of \(1.34\%\) on the \(1000^3\) GEMM. An unsound precondition (\(P\le S-1\)) allows an unsafe design that the solver then selects and the verifier catches.}
\label{fig:precondition}
\end{figure}
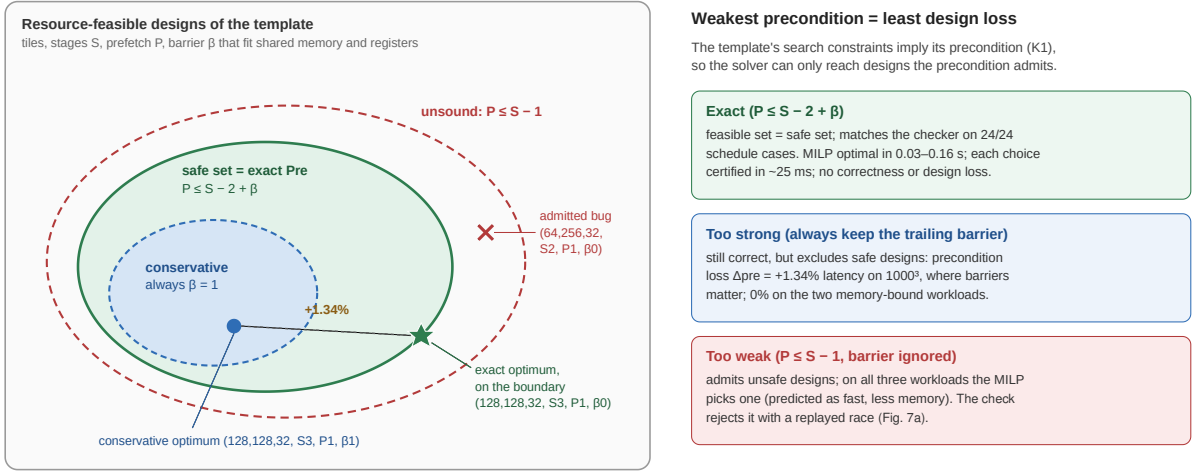

The cheap check of Section~\ref{checking-contracts-the-two-tier-trust-boundary} trusts the preconditions completely, and the preconditions decide what the solver may choose. Their strength therefore matters for both correctness and performance. Let \(\mathrm{Safe}\) be the set of assignments that really preserve the coarse behavior: the weakest precondition in Dijkstra's sense. A precondition stronger than \(\mathrm{Safe}\) is sound but throws away correct designs, and the cost is the \emph{precondition loss} \(\Delta_{\mathrm{pre}}=J^\star(\mathrm{Pre})-J^\star(\mathrm{Safe})\ge0\). A precondition weaker than \(\mathrm{Safe}\) allows incorrect designs, and the solver may pick one. Verification wants the honest boundary of correctness, and optimization wants the largest correct search space; both want the same set. Figure~\ref{fig:precondition} shows both failure modes for the shared-memory tile of Listing~\ref{lst:tiling-contract}. The precondition \(P\le S-2+\beta\), exact on the 24 checked combinations, contains the solver's optimum on its boundary. Always keeping the trailing barrier is sound but too strong, and costs 1.34\% of latency on a ragged GEMM. Dropping the barrier from the precondition is too weak: the solver then selects a design with a race, which the verifier rejects with the replayed race of Figure~\ref{fig:race} (Appendix~\ref{app:contract-details}).

\hypertarget{incremental-reuse-keeping-the-gate-affordable}{%
\subsection{Incremental reuse}\label{incremental-reuse-keeping-the-gate-affordable}}

The last piece is speed over time. The formulation changes at every step and at every repair, and rechecking everything after each change would make verification the bottleneck. Step-by-step refinement makes reuse natural: a new step adds a few conditions, and the proofs of the steps above it rest on premises that did not change.

The flow keeps its proofs in a directed acyclic dependency graph whose nodes are definitions, assumptions, statements and checked proofs. Constraints are kept as named groups, so a lemma depends only on the groups it uses. After a change, the nodes whose content changed and every node that depends on them, directly or through others, are rechecked in topological order; everything else is kept. Because the graph has no cycles and nodes are rechecked after all their premises, no proof is ever justified by itself or by a premise that has not been rechecked.

\refstepcounter{paperTheorem}\label{thm:incremental-reuse}\textbf{Theorem \thepaperTheorem (soundness of incremental reuse).} Suppose the dependency graph is acyclic, dependency information is complete, all affected nodes are rechecked against their revised statements and contexts, and acceptance requires accepted non-root premises. Then every retained or newly accepted theorem has a valid derivation under the revised root assumptions.

In plain words, a kept proof is safe as long as nothing it depends on has changed. A common case is a repair that only adds constraints, for instance when the agent tightens a template after a counterexample.

\refstepcounter{paperCorollary}\label{cor:constraint-strengthening}\textbf{Corollary \thepaperCorollary (reuse under constraint strengthening).} If parameter, design and input domains, the workload contract, execution semantics and environment assumptions are unchanged, and \(C_{\mathrm{new}}\Rightarrow C_{\mathrm{old}}\), then soundness of the old formulation implies soundness of the new one.

For the five-step GEMM chain, reuse cuts the solver queries at depth five from 92 to 27 and speeds up checking by 2.51$\times$ (Section~\ref{sec:exp-verification}), and the saving grows overall with refinement depth. That is the case where new steps sit on unchanged ones. When the agent edits an existing program, less carries over, and the measured saving is 1.25$\times$.

Together, local proofs, one-time template proofs, weakest preconditions and reuse are what let verification run inside the loop. Where every step carries a proof, the flow's output is a design, the bounds that support it, and a proof that it computes the workload. Section~\ref{evaluation} tests each part.

% Full-paper entry. Load experiments/latex/format.tex in the host preamble.
% Replace the host's Evaluation body with this input; do not append a duplicate.
\section{Evaluation}\label{evaluation}
\begingroup
\renewcommand{\ExperimentRoot}{experiments/latex}
\let\ExperimentSection\subsection
\let\ExperimentSubsection\subsubsection
\let\ExperimentSubsubsection\paragraph
% Shared experiment text. No document class, title, appendix switch, or counter reset.
The evaluation asks three questions, in the order of the paper's claims. First, on a fixed chip, are the templates the agent writes as good as production software? We run complete layers of two production models on NVIDIA GB10 against SGLang (Section~\ref{sec:exp-software}), compare the flow with a black-box coding agent given the same budget (Section~\ref{sec:construction-control}), and test both fixed-template reuse and knowledge-assisted construction on new tasks (Section~\ref{sec:exp-efficiency}). Second, does letting the chip change with the software beat software alone? We compare joint design with software-only, hardware-only and sequential design on the Ventus GPU under a fixed area budget (Section~\ref{sec:exp-codesign}). We then test whether wall relaxation reaches the joint solution by purchasing capacity that enables better software, and compare an explicit Gemmini formulation with DoSA under its energy--delay objective. Third, is verification cheap enough to run inside the loop, and does it catch real faults? We measure proof reuse and fault detection on the five-step GEMM chain and during actual compiled-program revisions (Section~\ref{sec:exp-verification}). The studies are separate, and each supplies a different kind of evidence with its own limits. Table~\ref{tab:evidence} lists, for each study, what it shows, who supplied which part, and what it does not show. Section~\ref{sec:exp-setup} first describes the two settings and how we measure.

\begin{table}[!htb]\centering\footnotesize
\caption{What each study shows and where it stops.}\label{tab:evidence}
\begin{tabularx}{\linewidth}{@{}>{\raggedright\arraybackslash}p{.15\linewidth}>{\raggedright\arraybackslash}X>{\raggedright\arraybackslash}X>{\raggedright\arraybackslash}X@{}}\toprule
Study & Evidence & Who supplied what & What it does not show\\\midrule
GB10 layers (\ref{sec:exp-software}) & Device latency of complete layers against SGLang; numerical checks on tested inputs & agent wrote program families and cost expressions; production components inherited; solver selects the instance & No formal proof of the GPU program families\\
White vs.\ black box (\ref{sec:construction-control}) & One construction run per method from the same eager code under equal caps & agent wrote both programs; host-side help in diagnosis & No demonstrated selection gain over copying; construction-run variability\\
Template reuse / construction (\ref{sec:exp-efficiency}) & Fixed-family transfer; three construction runs per knowledge condition & Fixed families transferred; agent constructs new families; solver selects instances & Total construction cost including inherited knowledge\\
Ventus co-design (\ref{sec:exp-codesign}) & Cycle simulation, calibrated aggregation and extrapolation & Study menus and references set by the authors; solver selects & Cross-study comparisons under a common reference\\
Wall relaxation (\ref{sec:exp-codesign}) & Exact inner solves on a 12-chip capacity domain; purchases confirmed by simulation & Author-defined menus; solver-based outer search & A general search-efficiency advantage; random search finds the same optimum\\
Gemmini (\ref{sec:exp-codesign}) & Analytical co-design on DoSA's BERT GEMM list with Timeloop/Accelergy & agent derived costs from the Gemmini source and wrote the enumeration & Automated template construction/reuse loop; complete Transformer inference\\
GEMM tiling (\ref{checking-contracts-the-two-tier-trust-boundary}--\ref{the-weakest-contract}, App.~\ref{app:contract-details}) & Z3 proofs and replayed counterexamples on 24 combinations & Template and bounded models written by hand & Schedules beyond two warps and five phases; FP32 rounding\\
Verification reuse (\ref{sec:exp-verification}) & Checker time with and without reuse & Chain written by hand; seven real agent proposals & End-to-end savings: generation and compilation excluded\\\bottomrule
\end{tabularx}
\end{table}

\ExperimentSection{Setup}\label{sec:exp-setup}
We evaluate the flow in two settings (Table~\ref{tab:settings}). On NVIDIA GB10 the chip is fixed, and the agent writes templates for complete layers of two production models. Each GB10 template holds a family of TileLang programs with open decisions, such as tile sizes, split counts and which engine to use, plus a cost expression; a MIP solver picks one program per scenario. These templates carry no proofs. Each candidate is instead checked numerically against a reference before it is timed. On Ventus, a parameterized RISC-V GPU, software and chip resources are chosen together under a fixed area budget.

\begin{table}[!htb]\centering\small
\caption{The two experimental settings. GPU timing, Ventus timing models and checker runs are reported separately.}\label{tab:settings}
\begin{tabularx}{\linewidth}{@{}l*{2}{>{\raggedright\arraybackslash}X}@{}}\toprule
 & Fixed-hardware software & Hardware--software co-design\\\midrule
Platform & NVIDIA GB10; TileLang 0.1.15\ExperimentCitation{tilelang} & Ventus, parameterized RISC-V GPU\\
Workload & Qwen3-8B and Qwen3.5-35B-A3B\ExperimentCitation{qwen3-model,qwen35-model}; BF16, batch one & Qwen2.5-0.5B\\
Input scenarios & Prefill 128/512/2048/4096; decode history 512/2048/4096/8192 & Prefill 128/512; 16 decode steps from history 128/512\\
Design freedom & Graph structure, kernels, layouts, schedules & Same software freedoms within the study menus, plus chip resources\\
Primary metric & Complete-scope device latency; service TTFT/TPOT & Worst-scenario normalized cycles at a fixed area budget\\
Efficiency & Recorded search/build cost and reuse controls & Formulation/search cost and incremental verification cost\\\bottomrule
\end{tabularx}
\end{table}

\lead{Protocol.}
GPU latencies are medians pooled over repeated blocks, and every comparison times all its methods in the same batch. Timing scopes differ and are stated with each result: dense layers are timed as complete layer calls over three paired blocks (Table~\ref{tab:software-final}); the MoE group of Table~\ref{tab:moe-final} and the attention studies use CUDA Graph replay; Figure~\ref{fig:moe-eager} times the MoE group without Graph capture, so that eager can be included; and serving reports time to first token and throughput. Every candidate must pass numerical output and state checks against a reference before it is timed; a candidate that fails has no reported speedup, shown as a dash. Packing, compilation and solve times are reported separately from latency. Speedups are computed before rounding the displayed latencies. P and D denote prefill length and decode history. We wrote the tiling contract and the bounded verification models by hand. The agent wrote the GPU programs and cost expressions.

The additional Gemmini study uses the BERT operator list and energy--delay objective published with DoSA. Its modeled energy and area are reported separately from Ventus cycle ratios. The attention construction study uses Triton on GB10 and three independent agent runs per knowledge condition; its implementation and timing protocol are specified with that comparison.

\ExperimentSection{Templates Proposed by the agent on GB10}\label{sec:exp-software}\label{sec:exp-kernels}\label{sec:kernels}
The agent writes program families and cost expressions for complete layers of Qwen3-8B and Qwen3.5-35B-A3B\ExperimentCitation{qwen3-model,qwen35-model}, and the solver selects one instance per scenario. We compare with SGLang's production path\ExperimentCitation{sglang,flashinfer} and with pure PyTorch eager. Figure~\ref{fig:complete-scopes} gives all sixteen scenarios.

\begin{figure}[!htb]\centering
\includegraphics[width=\linewidth]{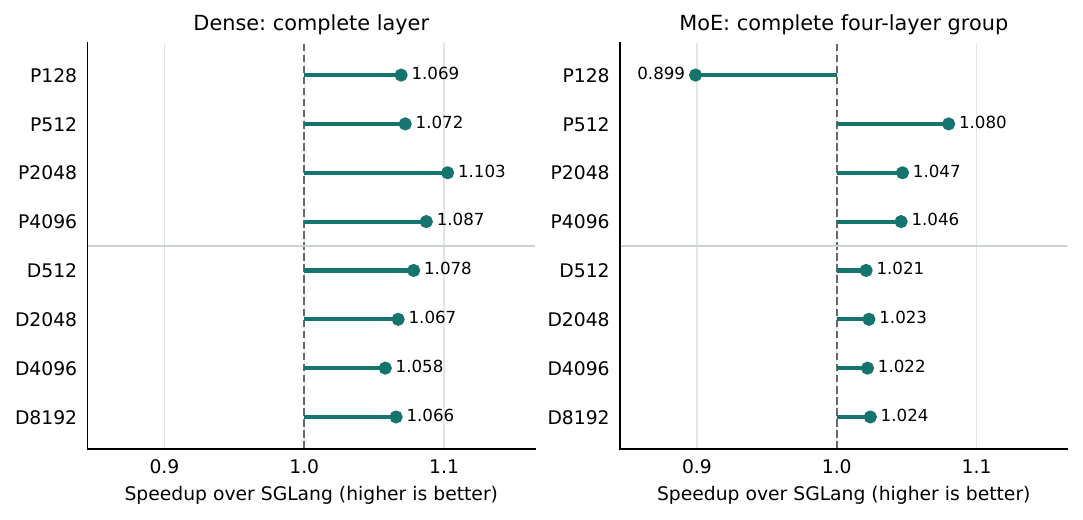}
\caption{Complete-scope performance across the eight scenarios of each model. Each point is the SGLang-to-Formulaic latency ratio reported in Tables~\ref{tab:software-final} and~\ref{tab:moe-final}; the dashed line denotes equal performance. P denotes prefill length and D denotes decode history. Both panels use the same scale.}\label{fig:complete-scopes}
\end{figure}

\lead{Dense layers.}
Table~\ref{tab:software-final} covers the complete Qwen3-8B layer: normalization, projections, attention, KV updates, MLP and residuals. Formulaic is faster than SGLang in all eight scenarios, with a 1.075$\times$ geometric mean, and 2.127$\times$ faster than eager. The candidate combines kernels the agent wrote with optimized production components it inherited; Section~\ref{sec:construction-control} evaluates construction from eager code alone. Timing each part separately shows where the gain comes from: the output and MLP block is 1.137$\times$ faster than SGLang at P2048, while the attention block matches production at prefill and gains 1.107$\times$ at D4096.

\begin{table}[!htb]\centering\small
\caption{Qwen3-8B complete-layer latency (\us), BF16, batch one. SGLang uses the pinned native-kernel path.}\label{tab:software-final}
\begin{tabular}{@{}lrrrrr@{}}\toprule
 & & & & \multicolumn{2}{c}{Speedup over}\\
Scenario & Eager & SGLang & Formulaic & Eager & SGLang\\\midrule
Prefill 128 & 2211.4 & 1920.7 & 1796.3 & $1.231\times$ & $1.069\times$\\
Prefill 512 & 4797.8 & 2851.8 & 2659.7 & $1.804\times$ & $1.072\times$\\
Prefill 2048 & 36476.3 & 11121.9 & 10087.9 & $3.616\times$ & $1.103\times$\\
Prefill 4096 & 153917.9 & 22799.9 & 20969.3 & $7.340\times$ & $1.087\times$\\
Decode 512 & 2255.1 & 1811.8 & 1680.3 & $1.342\times$ & $1.078\times$\\
Decode 2048 & 2588.5 & 1819.9 & 1705.4 & $1.518\times$ & $1.067\times$\\
Decode 4096 & 2979.2 & 1839.4 & 1738.7 & $1.713\times$ & $1.058\times$\\
Decode 8192 & 3732.5 & 1953.2 & 1832.8 & $2.037\times$ & $1.066\times$\\
\midrule
Geomean & -- & -- & -- & $2.127\times$ & $1.075\times$\\\bottomrule
\end{tabular}
\tnote{SGLang 0.5.9 with FlashInfer is the fastest production path we could run in every row. All eight eager comparisons pass their numerical checks. Each row compares paths measured in the same batch. Decode query length is one.}\end{table}

\lead{Hybrid attention and MoE.}
Qwen3.5-35B-A3B alternates three Gated DeltaNet layers with one full-attention layer, each with top-8 routing over 256 experts and a shared expert. The optimization targets expert execution and keeps the production attention mixers. Over the complete four-layer group (Table~\ref{tab:moe-final}), seven of eight scenarios improve, with a 1.019$\times$ geometric mean; P128 regresses to 0.899$\times$. Against pure PyTorch eager, the same group reaches 5.42--6.84$\times$ across four decode histories with implementations that pass the output and state checks (Figure~\ref{fig:moe-eager}), from 17.97--23.14~ms to 3.27--3.55~ms. At D2048, matching the projection accumulation structure removes a rounding-induced expert-routing divergence; the repair uses custom Triton kernels and passes the original output and state checks. This implementation repair is measured separately from the MIP-selected CUDA Graph results in Table~\ref{tab:moe-final}. Much of this gain over eager comes from the production attention mixers that the candidate keeps; the agent's own part, the expert path, is measured against SGLang in Table~\ref{tab:moe-final}.

\begin{table}[!htb]\centering\small
\caption{Qwen3.5-35B-A3B complete four-layer group latency (\us), BF16, batch one. Each group contains three linear-attention layers and one full-attention layer, all with MoE.}\label{tab:moe-final}
\begin{tabular}{@{}lrrr@{}}\toprule
Scenario & SGLang & Formulaic & Speedup\\\midrule
Prefill 128 & 22411.5 & 24938.8 & $0.899\times$\\
Prefill 512 & 31366.2 & 29034.8 & $1.080\times$\\
Prefill 2048 & 53139.5 & 50752.8 & $1.047\times$\\
Prefill 4096 & 83451.5 & 79748.2 & $1.046\times$\\
Decode 512 & 2843.7 & 2784.0 & $1.021\times$\\
Decode 2048 & 2873.9 & 2808.5 & $1.023\times$\\
Decode 4096 & 2922.9 & 2860.8 & $1.022\times$\\
Decode 8192 & 3009.9 & 2938.4 & $1.024\times$\\
\midrule
Geomean & -- & -- & $1.019\times$\\\bottomrule
\end{tabular}
\tnote{First four layers, including routing, state and KV updates, normalization and residuals; CUDA Graph replay. The production control is the faster of SGLang's serial and dual-stream paths for each scenario.}\end{table}

\begin{figure}[!htb]\centering
\includegraphics[width=.95\linewidth]{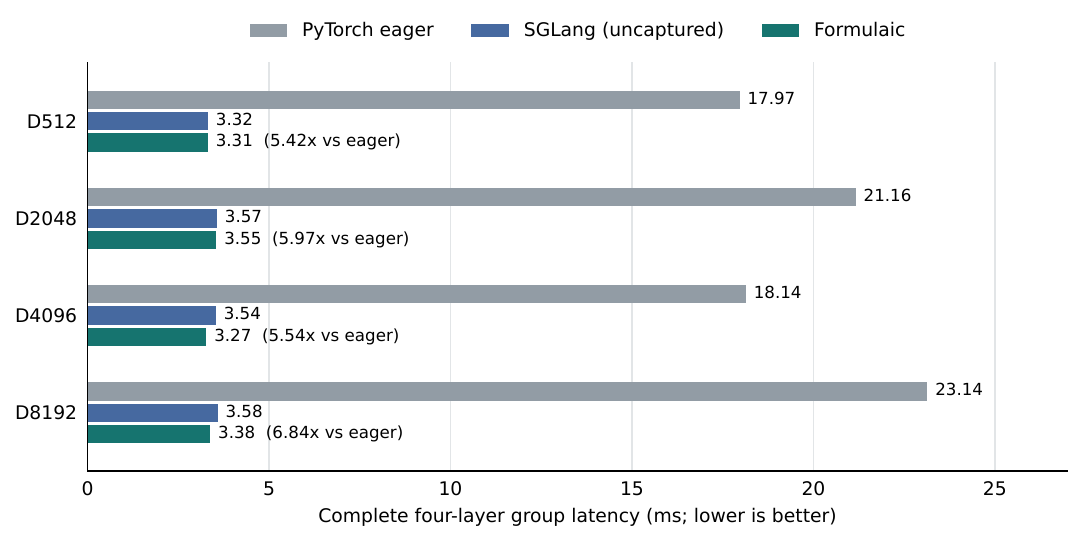}
\caption{Complete MoE four-layer decode group, measured without CUDA Graph capture. Each row compares all three paths in one timing batch. D2048 uses repaired MMA-based expert projections and newly measured controls; other rows retain their independently measured implementations. The repaired D2048 candidate achieves 5.97$\times$ speedup over eager and approximately matches SGLang.}\label{fig:moe-eager}
\end{figure}

\lead{Serving.}
In end-to-end serving with 32 output tokens, replacing prefill lowers dense time to first token by 5.72\% and 4.73\% at prompts of 2048 and 4096 tokens, but raises it by 5.82\% at 512; MoE prefill replacement across all 40 layers lowers it by about 6.3\% at prompts of 512 and 2048 tokens. Output tokens are identical to production in these runs. Replacing dense decode as well raises single-client throughput at P2048 by 7.70\%. Two results limit the claim. The P512 slowdown shows that a gain on one hot layer need not carry over to serving. And at full-model scope the MoE decode replacement fails a 40-layer output and state check, so serving uses it for prefill only.

\ExperimentSection{White Box versus Black Box}\label{sec:construction-control}
This study asks whether formulating helps, compared with an agent that writes kernels directly. Both runs start from the same PyTorch eager code for Q1 grouped-query attention, with the same hardware information, model, and caps on agent generation and device time; both may revise source. The direct agent searches by writing and measuring code. Formulaic also writes a cost expression and lets the MIP select the instance. Programs are fixed before the validation measurements.

\begin{figure}[!htb]\centering
\includegraphics[width=\linewidth]{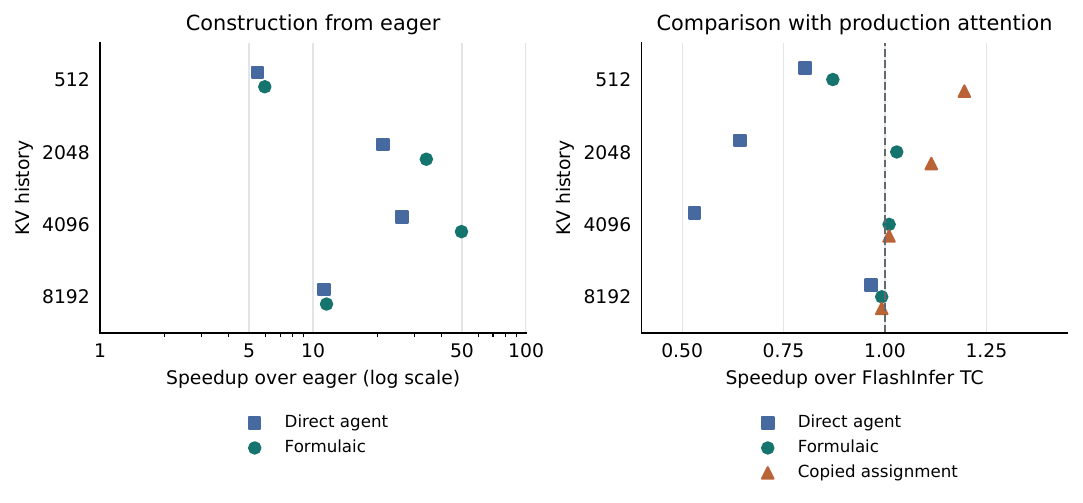}
\caption{Matched-budget construction across all four histories. Left: speedup over eager on a logarithmic scale. Right: speedup over the faster FlashInfer mode (TC), including the copied development assignment. Values are ratios of the same-row pooled medians in Table~\ref{tab:efficiency}. Higher is better in both panels.}\label{fig:construction-comparison}
\end{figure}

\begin{table}[!htb]\centering\small
\caption{Programs built from eager code and fixed before measurement, across four histories, in \us. Each row measures all paths in one process after warming each implementation immediately before its timing block. Formulaic uses the MIP-selected instance; copying the development assignment is reported separately. This compares attained performance, not total optimization cost.}\label{tab:efficiency}
\begin{tabular}{@{}rrrrrrr@{}}\toprule
History & Eager & Direct & Formulaic & Copied & FI TC & FI SIMT\\\midrule
512 & 46.14 & 8.44 & 7.78 & 5.67 & 6.78 & 12.02\\
2048 & 374.22 & 17.61 & 11.00 & 10.16 & 11.32 & 15.56\\
4096 & 783.15 & 30.01 & 15.74 & 15.74 & 15.90 & 20.34\\
8192 & 1597.65 & 141.98 & 138.16 & 138.16 & 137.05 & 138.41\\\bottomrule
\end{tabular}
\tnote{Entries are medians pooled across blocks for the primary implementation copy; a second copy of the same implementation measures repeatability. FI denotes FlashInfer; TC uses tensor cores and SIMT uses single-instruction, multiple-thread execution, corresponding to \texttt{use\_tensor\_cores=True/False}; TC is faster for the primary copy in all four rows and fixed for paired comparisons. All paths pass finite FP32-reference, CUDA Graph, input-immutability and post-timing checks. Histories 512 and 8192 are new validation inputs; 2048 repeats the protocol and 4096 was used in development. The program and cost model are fixed before these measurements.}
\end{table}

Both runs used GPT-6 Astra at medium reasoning effort, with caps of 1200~s of agent generation (at most 600~s per call) and 3300~s of device jobs per run, and both received host-side help in diagnosis. Formulaic builds a grouped-head BF16 tensor-core branch and exposes engine, tile, split and warp decisions to the solver. Its selected programs run 5.93--49.76$\times$ faster than eager, with lower measured median latency than the direct agent at all four histories: the gap ranges from 2.8\% at 8192 to 1.9$\times$ at 4096 (Table~\ref{tab:efficiency}). At 4096 it matches the faster FlashInfer mode within measurement variation. It also used less agent-generation time, 611~s against 974~s; the latter includes a cancelled 69~s call. The gain comes from the program family the agent built, not from the solver's choice of instance. Copying the assignment found during development at history 4096 is faster than the solver's choice at 512 and 2048 and equal at 4096 and 8192, so this study isolates no gain from solver selection. The cost model also underpredicts the 8192 latency by 2.92$\times$. Each method was run once, so this shows what each can reach, not that one is always better.

\ExperimentSection{Reuse of Templates}\label{sec:exp-efficiency}
We test two uses of accumulated templates: re-solving an unchanged family for a new workload, and using prior knowledge to construct a new family.

\lead{Transferring a fixed family.}
With source and expression unchanged, the GateUp template runs 1.067--1.319$\times$ faster than SGLang on Qwen2.5-7B weights at four input sizes. Each new shape requires about 17--19 seconds of solving and 3 seconds of building and packing. A fixed attention template likewise transfers to four new history lengths, attaining 1.9--3.6$\times$ speedup over a tiled online-softmax attention baseline. Re-solving can improve the chosen instance, although at one length the expression misses a faster split. The full transfer tables and selection controls are in Appendix~\ref{app:reuse-details}.

\lead{Using knowledge to construct a new family.}
We compare construction from a common task specification with construction supplied with an executable template, its cost expression and explanatory optimization knowledge. Each condition has three independent agent runs on each of three single-query attention tasks, giving 18 trajectories in this comparison. Runs receive the same within-task budgets and may revise both the program family and expression using their own feedback. Final expression-selected programs are checked numerically and timed together with FlashInfer on GB10 using warmed CUDA Graph replay. Task shapes, budgets, latency ranges and costs appear in Appendix~\ref{app:reuse-details}.

\begin{table}[!htb]\centering\small
\caption{Does prior knowledge help the agent construct a faster attention program? Entries are speedups over FlashInfer, using median latency across three independent runs per cell. Values above 1 beat FlashInfer; below 1 are slower. All conditions use the same tools and within-task budgets.}\label{tab:knowledge-summary}
\begin{tabular}{@{}rrr@{}}\toprule
KV length & Task only & + Executable template\\\midrule
1025 &0.835$\times$&1.488$\times$\\
6145 &0.942$\times$&1.158$\times$\\
24577 &1.038$\times$&1.031$\times$\\\bottomrule
\end{tabular}
\end{table}

Prior knowledge improves the selected programs most clearly on the smaller two tasks (Table~\ref{tab:knowledge-summary}). At length 1025, two template runs introduce 17-way splitting and revise the merge and expression, avoiding an extra tile iteration incurred by 16 partitions. The template condition reaches 1.49$\times$ FlashInfer and 8.98$\times$ eager performance. The template condition also improves performance at length 6145, while the two conditions are nearly tied at length 24577. Thus the supplied family and knowledge help construction on two tasks, with comparable measured construction costs; the largest task shows no additional performance benefit.

\ExperimentSection{Hardware--Software Co-design}\label{sec:exp-codesign}
The Ventus evaluation\ExperimentCitation{r89} isolates the value of allowing software and hardware decisions to interact. Software-only optimization fixes the reference chip; hardware-only fixes the reference software. Formulaic's joint formulation releases both sets of variables. All strategies use the same final menus and workload objectives.

The compiled-kernel and queue-reallocation study uses Qwen2.5-0.5B, batch one, BF16 storage and FP32 accumulation, within a 1.11~mm$^2$ area budget. The reference has two SMs, eight resident 32-thread warps per SM, 128~KiB shared memory and 64~KiB L1 per SM, and 128~KiB shared L2. The hardware menu spans 27 fields. External-memory service stays at 64 bytes/cycle. Area uses synthesis-calibrated logic and modeled SRAM costs\ExperimentCitation{r92,r56}.

\begin{table}[!htb]\centering\small
\caption{Qwen2.5-0.5B co-design on Ventus at a 1.11~mm$^2$ area ceiling. The first three columns give cycles normalized to the reference software and chip (lower is better). The last column gives joint speedup over software-only optimization (higher is better).}\label{tab:codesign}
\begin{tabular}{@{}lrrrr@{}}\toprule
Objective & SW only & HW only & Joint & Speedup over SW\\\midrule
Prefill &0.758&0.440&0.377&2.01$\times$\\
Decode &0.934&0.454&0.428&2.18$\times$\\
All scenarios &0.934&0.454&0.428&2.18$\times$\\\bottomrule
\end{tabular}
\tnote{The queue study simulates 128-token prefill and extrapolates the 512-token scenarios.}
\end{table}

Table~\ref{tab:codesign} evaluates compiled-kernel changes and queue reallocation; the following study reopens tile size, register capacity, tensor shape and cache allocation. Each study normalizes cycles to its own reference chip and software.

\lead{Changing both sides exposes gains unavailable to either side alone.}
In complete P128 simulation, the compiled-kernel and queue combination reduces execution from 812.6 to 264.1 million cycles, compared with 539.8 million for the software change alone, at 1.108~mm$^2$. The software reduces indexing and access overhead; queue reallocation absorbs the remaining requests. Table~\ref{tab:codesign} shows the combination beating both single-side controls.

A complementary complete instruction-simulation study adds a 64-by-64 tile requiring 173 vector registers. It is unfavorable on the earlier chip; reopening register capacity, tensor shape, and cache allocation makes it useful. The balanced design achieves cycle ratios of 0.342/0.331 for P128/P512 and 0.295/0.272 for D128/D512 at 1.081~mm$^2$, giving 2.93$\times$ worst-scenario speedup. A decode-specialized design reaches 3.58$\times$ on decode while sacrificing prefill. The objective determines which coupling is valuable.

The same principle applies to constrained storage: jointly choosing the KV representation and hardware for one single-layer Qwen2.5-0.5B decode step reduces storage by 16--17\% relative to software-only optimization, at latency ratios near 0.18--0.20 and local attention-error thresholds of 2\% or 5\%. These studies show the value of joint design for their stated objectives and simulators. Appendix~\ref{app:model-fidelity} reports the cost-model and chip-transfer controls.

\lead{Wall relaxation recovers the joint solution by purchasing capacity.}
We directly evaluate the algorithm on Qwen2.5-0.5B with prefill, decode and mixed objectives at lengths 128 and 512. Nine prefill and seven decode programs are searched over twelve vector/scalar register-capacity combinations under a 1.11~mm$^2$ ceiling. Other hardware fields are fixed, and each admitted kernel fits eight resident one-warp blocks, so capacity changes enable different programs without changing the cost of an admitted program. All methods share this domain and fresh instruction-simulation coefficients. The objective is the worst scenario-specific cycle ratio to a common reference.

From the smaller starting chips, wall relaxation improves prefill, decode and mixed performance by 2.77$\times$, 1.60$\times$ and 2.38$\times$, respectively, reaching the joint optimum within this finite domain. Prefill first buys scalar and then vector capacity, enabling faster programs. Instruction simulation confirms all ten purchase gains predicted from the simulator-derived coefficients. Random search with a twelve-chip budget reaches the same optimum, so this experiment demonstrates capacity-driven software reselection without establishing a search-cost advantage. Table~\ref{tab:wall-capacity} gives both starting chips and all methods, and Figure~\ref{fig:wall-capacity-path} plots quality against search work. From the larger starting chips, the start is already optimal, so wall relaxation's three inner solves only check this and stop. Sequential design finds nothing from either start. In this capacity-only model a program's cost does not depend on the chip, so with the software fixed, buying capacity cannot help; this limits what the comparison with sequential design tests. Purchase validation follows search; online model repair is not evaluated here. Appendix~\ref{app:wall-details} gives the full protocol and validation costs.

\begin{table}[!htb]\centering\small
\caption{Capacity search across three objectives and two starting chips. Each cell gives the final simulated cycle ratio, followed by the number of distinct inner problems in parentheses; lower is better. Column headers specify vector/scalar register slots per SM.}\label{tab:wall-capacity}
\begin{tabular}{@{}lrrrrrr@{}}\toprule
 & \multicolumn{2}{c}{Prefill} & \multicolumn{2}{c}{Decode} & \multicolumn{2}{c}{Mixed}\\
\cmidrule(lr){2-3}\cmidrule(lr){4-5}\cmidrule(l){6-7}
Method &1024/128&2048/256&512/256&2048/256&512/256&2048/256\\\midrule
Fixed hardware &.642 (1)&.231 (1)&.431 (1)&.269 (1)&.642 (1)&.269 (1)\\
Sequential &.642 (1)&.231 (1)&.431 (1)&.269 (1)&.642 (1)&.269 (1)\\
Random + exact inner &.231 (12)&.231 (12)&.269 (12)&.269 (12)&.269 (12)&.269 (12)\\
Wall relaxation &.231 (12)&.231 (3)&.269 (8)&.269 (3)&.269 (11)&.269 (3)\\
Coupled wall &.231 (16)&.231 (4)&.269 (11)&.269 (4)&.269 (16)&.269 (4)\\
Joint reference &.231 (12)&.231 (12)&.269 (12)&.269 (12)&.269 (12)&.269 (12)\\\bottomrule
\end{tabular}
\tnote{Queries distinguish the chip and relaxed-constraint set, including virtual capacity relaxations. Sequential search also checks twelve hardware assignments with its fixed software. Random uses five seeds and a fixed twelve-chip budget; all seeds finish at the same quality. The joint reference is shared between starts and is exact within the declared domain.}
\end{table}

\begin{figure}[!htb]\centering
\includegraphics[width=\linewidth]{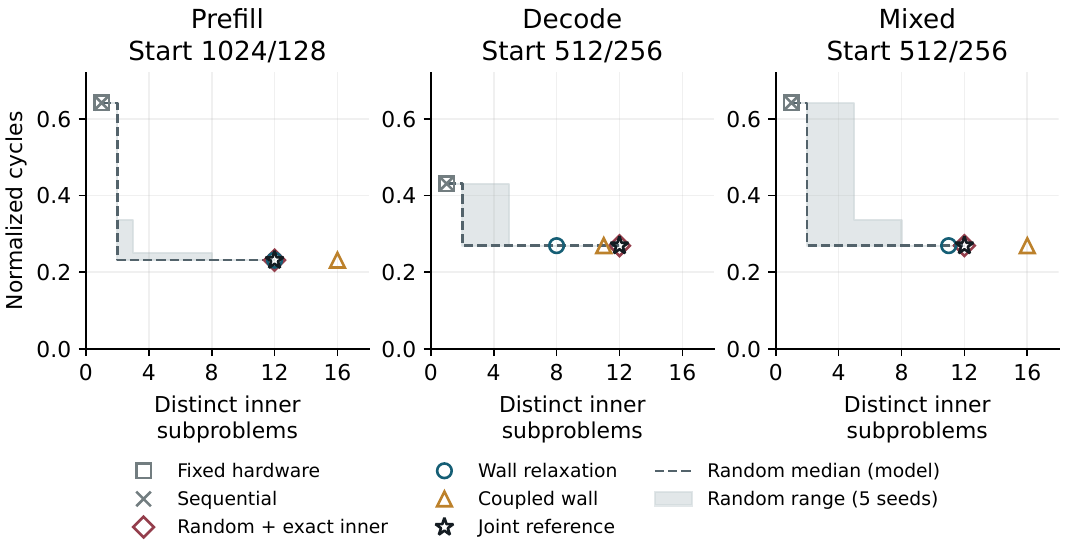}
\caption{Quality and search work from the smaller starts. Markers show final selected points checked by instruction simulation. The dashed curve and shaded range show the frozen-model random-search trajectories across five seeds; intermediate random points are not independently replayed. Overlapping markers have the same quality and query count.}\label{fig:wall-capacity-path}
\end{figure}

\lead{Explicit formulation also improves an external accelerator objective.}
On DoSA's published BERT GEMM list\ExperimentCitation{r10} and common Gemmini evaluator, a source-informed formulation reduces energy--delay product by 16.53\% and estimated area by 28.71\% relative to the complete seven-start DoSA run, at equal latency. This comparison follows DoSA's energy--delay objective without an area ceiling, complementing the constrained Ventus studies. Table~\ref{tab:dosa-bert} reports the comparison. The workload has five distinct GEMMs with six weighted occurrences. DoSA uses seven starts of 1490 updates in one run with random seed 0. Our agent read the Gemmini model source, derived access and execution costs from it, enumerated mapping choices, and combined layer plans across 1008 hardware-budget points. This result supports the effectiveness of agent-built formulations for co-design. A fully automated loop for independent template construction and reuse has yet to be evaluated. Seven proposals are evaluated using the same minimum-hardware construction and Timeloop/Accelergy policy. Both searches limit array dimensions to 128 and use 1~KiB memory granularity; tiny memories share a capacity-preserving register-file model without calibrated layout costs.

\begin{table}[!htb]\centering\small
\caption{BERT operator-list optimization on Gemmini. Lower energy, latency, energy--delay product (EDP) and area are better. All rows use the same Timeloop/Accelergy evaluation policy.}\label{tab:dosa-bert}
\begin{tabular}{@{}lrrrr@{}}\toprule
Method & Energy ($\mu$J) & Latency (cycles) & EDP ($10^9\,\mu$J cycles) & Area (mm$^2$)\\\midrule
Default chip + CoSA &2631.85&2,963,456&7.7994&0.708\\
DoSA &1983.04&917,508&1.8195&14.088\\
Formulaic &1655.24&917,508&1.5187&10.043\\\bottomrule
\end{tabular}
\tnote{The workload is the published weighted GEMM list, not complete Transformer inference. DoSA selects a 128-by-128 array with 896~KiB scratchpad and 512~KiB accumulator; Formulaic selects the same array dimension with 192~KiB and 384~KiB, respectively. The default chip uses a 16-by-16 array, 128~KiB scratchpad and 32~KiB accumulator.}
\end{table}

\ExperimentSection{Verification Cost and Fault Detection}\label{sec:exp-verification}
The five-step GEMM chain compares rechecking every earlier step with reusing a proof when its premises are unchanged. Both modes disable individual-query caching and reject faults injected at every step. At depth five, progressive verification reduces queries from 92 to 27 and achieves a paired 2.51$\times$ checker speedup. This is the saving when each new step sits on unchanged ones; when the agent edits a program, less carries over, as the next study shows.

\begin{table}[!htb]\centering\small
\caption{Measured verification cost along the refinement chain. Times are medians of seven paired repetitions; speedups are paired statistics. Progressive is our proof-reuse strategy.}\label{tab:verification-chain}
\begin{tabular}{@{}lrrrr@{}}\toprule
Added refinement & Full (ms) & Progressive (ms) & Queries, full/ours & Speedup\\\midrule
Reduction partition & 25.5 & 26.3 & 8/8 & 0.97$\times$\\
Storage & 73.8 & 48.7 & 23/15 & 1.51$\times$\\
Asynchronous transfer & 289.0 & 200.6 & 43/20 & 1.50$\times$\\
Double buffering & 528.1 & 218.9 & 65/22 & 2.41$\times$\\
Shared resources & 958.4 & 380.6 & 92/27 & 2.51$\times$\\\bottomrule
\end{tabular}
\end{table}

\begin{samepage}
\lead{Reuse remains useful when optimization changes the program.}
We compare full checking, reuse of unchanged proof-dependency nodes, and additional reuse of identical SMT queries on controlled revisions and an actual agent optimization trajectory. All strategies start with empty caches and charge identity checks, evidence authentication and maintenance. Controlled revisions across one to eight shapes achieve 1.59--1.72$\times$ speedup through node reuse. On real program changes, a large semantic node is invalidated, but identical obligations within it can still be reused (Table~\ref{tab:revision-summary}).\par
\end{samepage}

\begin{table}[!htb]\centering\small
\caption{Verification replay on an actual optimization history, in seconds. Seven agent proposals have three paired replays; the 59-version history includes 52 host diagnostic executions and has one replay. Speedups use paired ratios.}\label{tab:revision-summary}
\begin{tabular}{@{}lrrrr@{}}\toprule
Sequence & Full & Node reuse & Query reuse & Full/Query\\\midrule
7 agent proposals &10.74&10.56&8.56&1.25$\times$\\
59-version history &83.04&81.66&72.17&1.15$\times$\\\bottomrule
\end{tabular}
\end{table}

Query reuse reduces checking time by 1.25$\times$ on the seven agent proposals and by 13.1\% over the complete history, with unchanged classifications. The checker proves 45 versions within its fixed-shape, mathematical-real scope and leaves 14 unresolved. The savings concern verification replay, excluding agent generation and compilation; the trajectory also informed query-reuse development. Appendix~\ref{app:revision-details} reports the controlled sequences, checking protocol and fault agreement.

\lead{Local checks detect faults introduced at each refinement.}
Across eight structural groups---from reduction partitioning and buffer reuse to resource scheduling and two connected GEMMs---the checker evaluates 78 variants, identifies 31 new faults and three deadlocks, and separates inherited faults, unfinished proofs and infeasible designs (Appendix~\ref{app:refinement-checks}). Workload models accept 100 valid schedules and replay all 16 injected fault families. The accepted tiling set equals $P\leq S-2+\beta$ on all 24 checked combinations. A stronger contract costs 1.34\% modeled latency on the ragged compute-bound shape; a weaker contract selects a rejected race. These proofs are on bounded models and are separate from the numerical GPU checks and Ventus timing.

\ifdefined\FullPaper Appendix~\ref{app:contract-details} gives the tiling-contract ablations and six injected faults; Appendix~\ref{app:refinement-checks} gives the structural and workload checks.\fi

\FloatBarrier

\endgroup

\hypertarget{related-work}{%
\section{Related Work}\label{related-work}}

\textbf{Optimization-based co-design and design-space exploration.} Design decisions once left to heuristics are now mathematical programs: loop transformations \cite{r28}, accelerator scheduling \cite{r9}, rematerialization and memory allocation \cite{r15,r16}, and high-level synthesis \cite{r87}. Joint search over mappings and hardware is established \cite{r17,r10,r18,r85,r86}, supported by generation, implementation and simulation infrastructure \cite{r19,r83,r84,r20} and by models and simulators \cite{r54,r55,r56,r57,r58,r59,r60}. Most search a space fixed in advance, with a cost model taken as given. Our Ventus model-fidelity study shows why simulation must also test the cost model (Appendix~\ref{app:model-fidelity}). We keep the solver, but an agent writes and revises the formulation, so the space itself can change and reach across levels of the stack.

\textbf{Shadow prices and outer search.} Shadow prices come from linear-programming duality, where the optimal dual of a capacity constraint prices one unit of it \cite{bertsimas1997}. Integer programs have no such dual, and post-optimality analysis must bound or re-solve \cite{geoffrion1977}. Wall relaxation prices by re-solving, which the small inner solve makes cheap, while its outer search uses the gain-per-cost principle studied in budgeted submodular optimization \cite{khuller1999,sviridenko2004}. Differentiable design-space exploration such as DoSA follows a continuous relaxation \cite{r10}; a smooth surrogate can guide search, but its selections must be evaluated on the discrete design domain.

\textbf{AI compilers, kernels and serving systems.} MLIR, TensorIR and TVM structure compilation across levels \cite{r21,r22,r79}; Ansor and Triton search and compile tiled schedules \cite{r23,r76}; Mirage, Axon and Exo search algebraic, scheduling and instruction-level transformations, sometimes checked against a reference \cite{r24,r25,r26,r27}. Optimized attention algorithms \cite{r77,r78} and serving systems \cite{r80,r81,r82} set the bar, and language models now generate kernels \cite{r73,r74}. Prism represents a family of tensor programs as a symbolic graph, verifies its equivalence to the input program before choosing parameter values, and then instantiates those parameters by profiling \cite{prism}; this is close in spirit to our templates, which also prove a family once and then choose an instance. These systems mostly optimize software for an existing chip, with validation ranging from testing to formal checks. We search the software and the chip together, and for the GEMM chain and the tiling template we prove the programs a template allows instead of only testing them.

\textbf{AI for optimization, formulation and design.} Language models with automatic evaluation discover algorithms and mathematical constructions \cite{r1,r2}, and evolving agent populations accumulate reusable mechanisms \cite{r5}. OptiMUS builds optimization models from natural-language descriptions \cite{r3}, and FormuEvo evolves mixed-integer formulations toward ones that solve faster \cite{r4}. LLM4MIP and LLM4Branch accelerate the solve \cite{r29,r32}. Learning also reaches compilers, architecture, floorplanning and RTL \cite{r30,r31,r6,r7,r8}. These directions build on sketching and counterexample-guided synthesis \cite{r64,r61,r63}, Benders decomposition \cite{r66}, and library and population learning \cite{r68,r70,r71}. STOP, G\"odel Agent and the Darwin G\"odel Machine let an agent modify itself \cite{zelikman2023stop,yin2024godel,zhang2025dgm}. Dr. RTL optimizes RTL with an agent that admits a rewrite only after sequential equivalence checking and keeps a reusable library of rewrite skills \cite{r8}. Existing systems already combine agents with evaluators, formal checks, and reusable knowledge. We investigate how explicit optimization formulations and their correctness conditions can serve as a common interface across these components and across hardware--software decisions.

\textbf{Verification of optimization models, compilers and hardware.} CvxLean, EquivaMap and SOVER check that optimization models and reformulations mean what they should \cite{r33,r13,r14}, FLARE proves that a MILP reformulation preserves the original problem with LLM-assisted theorem proving \cite{r11}, and VIPR certifies solver results \cite{r35}. Design verification rests on weakest preconditions and refinement mappings \cite{r12,r37}; CompCert and Alive2 prove that transformations preserve meaning \cite{r53,r40}, TensorRight verifies tensor rewrites \cite{r43}, ILA, Bolt and Kami address the hardware--software interface \cite{r45,r47,r48}, and proof caching and retrieval-augmented proving make repeated proof practical \cite{r50,r52}. These verify models, reformulations, transformations and hardware, some through a whole compiler. Our question is related but different: whether \emph{every} design a formulation allows computes what the workload requires, checked cheaply inside an optimization loop at each level of a hardware--software stack. We show this for one GEMM chain and one template (Section~\ref{verification-certifying-each-refinement}).

\textbf{Where this paper sits.} These directions are converging. Formulaic is a first step toward combining them: an agent that writes cross-layer formulations, solvers that search them with bounds, a search over the chip that follows the software's binding limits, simulation that tests the formulation's assumptions, and verification that proves the designs a formulation allows. In this paper we test these pieces in separate studies.

\hypertarget{discussion-and-limitations}{%
\section{Conclusions}\label{discussion-and-limitations}}

This paper argued that some co-design gains are lost in the formulation, before any optimizer runs, and explored a division of work in which a language-model agent writes formulations, solvers optimize within them, and verification and measurement check them. In our flow, the agent refines the problem step by step from the computation graph toward the chip. Each step is a template, a family of programs with the conditions under which they are correct. Wall relaxation extends the same idea to the chip: it prices the hardware limits that bind the software and buys the steps worth their area.

The separate studies teach three things. First, on a fixed chip, templates the agent wrote can match production software: complete Qwen3-8B layers on NVIDIA GB10 run 1.075$\times$ faster than SGLang, and the agent's program family beat a black-box coding agent at every input length in one run each, although that gain came from the program family, not from the solver's choice of instance. Second, letting the chip change pays: in a simulated Ventus study, joint design cuts worst-scenario cycles to 0.428 of the reference, against 0.934 for software alone. Third, local proofs can be reused: on a five-step refinement chain checking is 2.51$\times$ faster, and on real agent edits 1.25$\times$. What these studies do not yet show is the whole loop running on one design, with agent-written templates that carry proofs, chosen by wall relaxation and checked by simulation. That integration is the main future work.

\textbf{Limitations.} The work has five limits. First, the capacity-search experiment validates wall relaxation on a finite software and hardware domain, where random search finds the same optimum; larger domains with coupled throughput changes remain to be studied. Its outer rule is greedy, its single-wall screen can miss relaxations that only pay jointly, its prices are only as good as the cost model's ranking, and a wall that no template states cannot be priced. Second, proofs cover only the GEMM chain and the tiling template, with schedules on bounded models and arithmetic treated symbolically, checked with Python and Z3; the GB10 programs are checked numerically on the tested inputs, not proved. Third, the comparison with a black-box agent has one run per method, with host-side help in diagnosis, so it shows what each approach can reach rather than how often. Fourth, the chip models are incomplete. The Ventus results use cycle simulation or calibrated aggregation with no power model. Cost-model error can change the selected design: zero-gap solutions predict cycle ratios near 0.333, while simulation gives 0.791--0.899. Qwen-selected chips also regress on GPT-2 and Pythia (Appendix~\ref{app:model-fidelity}). Fifth, the MoE decode path fails a 40-layer output and state check at full-model scope, so serving uses the MoE replacement for prefill only.

\textbf{Future work.} The limits point to the next steps. The first is to run the studies as one loop on one design. The second is to extend wall relaxation beyond the measured capacity domain on Ventus, including throughput-changing resources and GB10 resource constraints, while measuring its gap to joint search, inner-solve cost and purchase accuracy. The third is to close the gap between proofs and measurements, by proving the generated GPU programs at the template level rather than checking them on tested inputs. The fourth is to make the cost model change with the search, so that a mismatch between prediction and measurement becomes a new wall or a new term rather than a wrong choice, and to add power and physical design to the chip model. The last is to let the template population grow over many problems and measure how much each new problem gains from what earlier ones left behind.

\paragraph{Implementation and data availability.} The flow, the experiment records and the scripts that produce each table and figure are kept in the authors' repository, and each reported number is traced to a recorded run. The main software versions are TileLang 0.1.15, SGLang 0.5.9 with FlashInfer, Transformers 4.57.1 for the MoE eager control, HiGHS 1.15.1 (highspy) and Z3 4.16.0.0 (z3-solver). The Ventus studies use the open-source Ventus GPGPU \cite{r89}.

\clearpage
\appendix
\section{The tiling contract: complete checks and ablations}\label{app:contract-details}

\textbf{Setup.} The shared-memory tiling template of Listing~\ref{lst:tiling-contract} is implemented end to end. Its search constraints form a mixed-integer linear program solved with HiGHS \cite{r91}. Its correctness conditions are proved with the Z3 SMT solver \cite{r90}, and every counterexample is replayed by an independent interpreter. The cost model belongs to an illustrative chip contract with 128 SMs, 1.8 TB/s of DRAM bandwidth, 80 TFLOP/s in FP32 and a 40-cycle barrier. Index conditions are proved for every tensor extent at each menu tile size, and schedule conditions for every interleaving of an event model with two warps and five reduction phases.

\textbf{The precondition is exact.} We ran the checker on all 24 combinations of slot count \(S\in\{1,2,3\}\), prefetch distance \(P\in\{0,\dots,3\}\) and trailing barrier \(\beta\). The set of schedules it proves safe equals the template's precondition \(P\le S-2+\beta\) exactly, with no extra designs allowed and none wrongly excluded. Every rejected combination fails the same condition, lifetime. Table~\ref{tab:contract-safe} lists the safe prefetch distances. The result turns a piece of GPU programming folklore, that an extra buffer can replace the trailing \texttt{\_\_syncthreads()}, into a precise, checked condition: a third slot buys one more step of prefetch or the removal of the barrier, but not both.

\begin{longtable}[]{@{}lll@{}}
\caption{Safe schedules of the shared-memory tiling template. The verifier's accepted set equals the template's precondition \(P\le S-2+\beta\) on all 24 combinations.}\label{tab:contract-safe}\\
\toprule\noalign{}
Slots \(S\) & Trailing barrier & Safe prefetch distances \(P\) \\
\midrule\noalign{}
\endfirsthead
\toprule\noalign{}
Slots \(S\) & Trailing barrier & Safe prefetch distances \(P\) \\
\midrule\noalign{}
\endhead
\bottomrule\noalign{}
\endlastfoot
1 & yes & 0 \\
1 & no & none \\
2 & yes & 0, 1 \\
2 & no & 0 \\
3 & yes & 0, 1, 2 \\
3 & no & 0, 1 \\
\end{longtable}

\textbf{Search and checking.} Table~\ref{tab:contract-selections} shows the designs the solver selects for three GEMM shapes. The MILP has 93 variables, 90 of them binary, and 248 constraints. HiGHS proves each selection optimal at zero gap in 0.03--0.16 s, and brute-force enumeration of the same 722 feasible designs reaches the same objective in every case. Checking a selected design (12 conditions) takes about 25 ms. All three selections lie on the boundary of the precondition: three slots, prefetch distance one and \emph{no} trailing barrier. The optimizer spends shared memory to remove a barrier, which is legal only because the precondition is the weakest one.
\begin{longtable}{@{}>{\raggedright\arraybackslash}p{.20\linewidth}>{\raggedright\arraybackslash}p{.30\linewidth}>{\raggedright\arraybackslash}p{.17\linewidth}>{\raggedright\arraybackslash}p{.20\linewidth}@{}}
\caption{Designs selected by the template MILP. Every row passes all 12 conditions; times are modeled.}\label{tab:contract-selections}\\
\toprule
GEMM & $(T_m,T_n,T_k,S,P,\beta)$ & Shared memory & Modeled time\\\midrule
\endfirsthead
\toprule
GEMM & $(T_m,T_n,T_k,S,P,\beta)$ & Shared memory & Modeled time\\\midrule
\endhead
\bottomrule\endlastfoot
$4096^3$ & $(128,256,16,3,1,0)$ & 72 KiB & 1826.9 $\mu$s, memory-bound\\
$64\times4096\times4096$ & $(64,64,16,3,1,0)$ & 24 KiB & 75.1 $\mu$s, memory-bound\\
$1000\times1000\times1000$, ragged tails & $(128,128,32,3,1,0)$ & 96 KiB, at capacity & 53.1 $\mu$s, compute-bound\\
\end{longtable}

\textbf{What a wrong contract costs.} Replacing the template's precondition with a stronger or a weaker one, and keeping everything else fixed, reproduces both failure modes of Section~\ref{the-weakest-contract} (Table~\ref{tab:contract-ablation}).

\begin{longtable}[]{@{}
  >{\raggedright\arraybackslash}p{(\columnwidth - 6\tabcolsep) * \real{0.4000}}
  >{\raggedright\arraybackslash}p{(\columnwidth - 6\tabcolsep) * \real{0.2000}}
  >{\raggedright\arraybackslash}p{(\columnwidth - 6\tabcolsep) * \real{0.2000}}
  >{\raggedright\arraybackslash}p{(\columnwidth - 6\tabcolsep) * \real{0.2000}}@{}}
\caption{Contract ablation. Percentages are precondition loss in modeled latency relative to the exact contract.}\label{tab:contract-ablation}\\
\toprule\noalign{}
\begin{minipage}[b]{\linewidth}\raggedright
Precondition given to the solver
\end{minipage} & \begin{minipage}[b]{\linewidth}\raggedright
\(4096^3\)
\end{minipage} & \begin{minipage}[b]{\linewidth}\raggedright
\(64\times4096^2\)
\end{minipage} & \begin{minipage}[b]{\linewidth}\raggedright
\(1000\times1000\times1000\)
\end{minipage} \\
\midrule\noalign{}
\endfirsthead
\toprule\noalign{}
\begin{minipage}[b]{\linewidth}\raggedright
Precondition given to the solver
\end{minipage} & \begin{minipage}[b]{\linewidth}\raggedright
\(4096^3\)
\end{minipage} & \begin{minipage}[b]{\linewidth}\raggedright
\(64\times4096^2\)
\end{minipage} & \begin{minipage}[b]{\linewidth}\raggedright
\(1000\times1000\times1000\)
\end{minipage} \\
\midrule\noalign{}
\endhead
\bottomrule\noalign{}
\endlastfoot
Exact, \(P\le S-2+\beta\) & optimal, passes & optimal, passes & optimal, passes \\
Too strong: always keep the trailing barrier & +0.0\%, passes & +0.0\%, passes & \textbf{+1.34\%}, passes \\
Too weak: \(P\le S-1\), barrier ignored & \textbf{rejected by verifier} & \textbf{rejected by verifier} & \textbf{rejected by verifier} \\
\end{longtable}

\begin{figure}[!htbp]
\centering
\includegraphics[width=1\textwidth,height=\textheight]{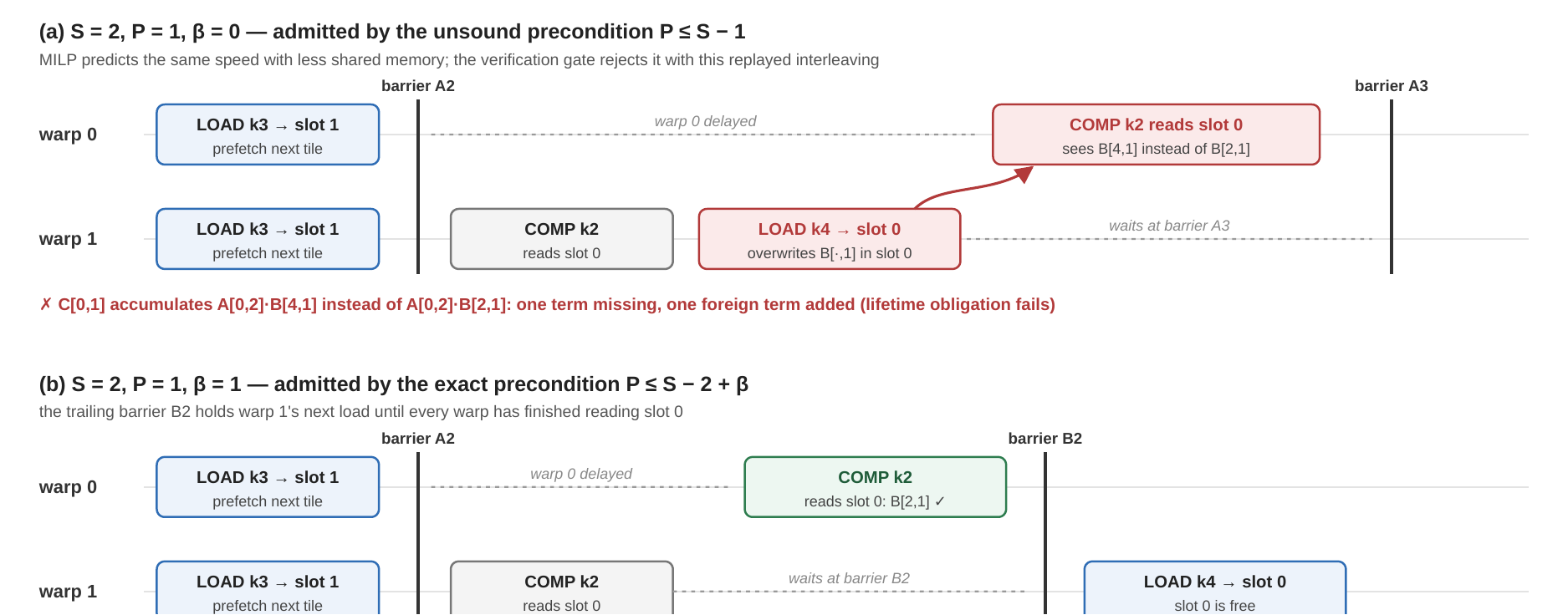}
\caption{The counterexample the verifier replays for the unsound precondition, with \(S=2\) slots and prefetch distance \(P=1\). (a) Without a trailing barrier (\(\beta=0\)), warp 1 finishes reading slot 0, crosses no barrier, and prefetches k-tile 4 into slot 0 while warp 0 is still waiting to read k-tile 2 from it. Output \(C[0,1]\) then accumulates \(A[0,2]\,B[4,1]\) instead of \(A[0,2]\,B[2,1]\): one term is lost and a foreign one is added. (b) With the trailing barrier (\(\beta=1\)), which the exact precondition \(P\le S-2+\beta\) requires for this design, the load waits until every warp has finished reading the slot.}
\label{fig:race}
\end{figure}

The stronger contract is safe but loses the optimum wherever barriers matter: 1.34\% of latency on the compute-bound ragged shape, and nothing on the memory-bound shapes, so precondition loss depends on the workload as well as on the template. The weaker contract makes the solver choose, on all three shapes, double buffering with prefetch one and no trailing barrier, which the cost model predicts to be just as fast while using less shared memory. The verifier rejects it and replays the race of Figure~\ref{fig:race}. The solver's answer is optimal for the constraints it was given, and wrong.

\textbf{Deliberately broken variants.} Starting from the proved \(1000\times1000\times1000\) design, we injected six faults that a plausible implementation could contain. The verifier rejects all six, each through the condition responsible and each with a replayed defect (Table~\ref{tab:contract-faults}). For tiles overlapping with stride \(T-1\), the coverage proof fails without a semantic witness, and the checker correctly records that failure as unresolved; the real defect is found by the exact-once condition that owns it.

\begin{longtable}[]{@{}
  >{\raggedright\arraybackslash}p{(\columnwidth - 4\tabcolsep) * \real{0.3636}}
  >{\raggedright\arraybackslash}p{(\columnwidth - 4\tabcolsep) * \real{0.2727}}
  >{\raggedright\arraybackslash}p{(\columnwidth - 4\tabcolsep) * \real{0.3636}}@{}}
\caption{Faults injected into the proved \(1000\times1000\times1000\) design, all rejected with a replayed defect.}\label{tab:contract-faults}\\
\toprule\noalign{}
\begin{minipage}[b]{\linewidth}\raggedright
Variant
\end{minipage} & \begin{minipage}[b]{\linewidth}\raggedright
Failing condition
\end{minipage} & \begin{minipage}[b]{\linewidth}\raggedright
Replayed defect
\end{minipage} \\
\midrule\noalign{}
\endfirsthead
\toprule\noalign{}
\begin{minipage}[b]{\linewidth}\raggedright
Variant
\end{minipage} & \begin{minipage}[b]{\linewidth}\raggedright
Failing condition
\end{minipage} & \begin{minipage}[b]{\linewidth}\raggedright
Replayed defect
\end{minipage} \\
\midrule\noalign{}
\endhead
\bottomrule\noalign{}
\endlastfoot
Floor-divided grid (tail uncovered) & coverage & missing reduction terms \\
Unguarded tail & no phantom & write outside \(C\); padding term added \\
Overlapping tiles (stride \(T-1\)) & exact-once & \(C[127,0]\) receives \(A[127,0]\,B[0,0]\) twice \\
Single buffer with prefetch one & lifetime & missing term \\
Double buffer, \(P=1\), no trailing barrier & lifetime & the race of Figure~\ref{fig:race} \\
Missing leading barrier & readiness, lifetime & reads of uninitialized slot elements \\
\end{longtable}

\textbf{Contract and instance cost.} For this template the search constraints and the correctness conditions coincide, so the solver's optimum is correct by construction and loses nothing to the contract. Soundness is argued once for the whole family; each instance then costs a solve of at most 0.16 s and a 25 ms check.

\FloatBarrier

\begingroup
\renewcommand{\ExperimentRoot}{experiments/latex}
\section{Template reuse and construction details}\label{app:reuse-details}

The fixed-family transfer measurements complement the construction comparison of Section~\ref{sec:exp-efficiency}.

\begin{table}[!htb]\centering\small
\caption{Measured reuse components for unchanged GateUp source and expression on Qwen2.5-7B layer-0 weights, input dimension $K=3584$ and intermediate dimension $I=18944$. Each row uses a separate paired comparison with production; activations are synthetic. Solver and build/packing columns are seconds, kernel columns microseconds. Construction of the original family is excluded.}\label{tab:reuse-cost}
\begin{tabular}{@{}rrrrrr@{}}\toprule
M & Solver total & Build/pack & SGLang & Reused selection & Speedup\\\midrule
128&19.060&2.878&1349.65&1264.97&1.067$\times$\\
512&19.162&2.851&1984.91&1505.20&1.319$\times$\\
2048&16.782&2.575&6980.16&5973.93&1.168$\times$\\
4096&17.081&2.499&13736.33&12096.62&1.136$\times$\\\bottomrule
\end{tabular}
\tnote{All 1536 feasible assignments per shape are checked. Static packing is 259~MiB and included in the build time. }
\end{table}

\begin{table}[!htb]\centering\small
\caption{An attention template reused on new history lengths. Each row has its own timing batch; the 1536 and 5120 rows check all six configurations. All implementations pass complete-output checks on the tested inputs.}\label{tab:reuse-heldout}
\begin{tabular}{@{}rrrrr@{}}\toprule
History KV & Split & Online (\us) & Formulaic (\us) & Paired speedup\\\midrule
3072 & 4 & 66.464 & 20.043 & 3.296--3.383$\times$\\
6144 & 4 & 131.915 & 67.573 & 1.942--2.200$\times$\\
\midrule
1536 & 4 & 34.882 & 11.442 & 3.047--3.052$\times$\\
5120 & 16 & 108.400 & 29.826 & 3.637--3.644$\times$\\
\bottomrule
\end{tabular}
\end{table}

Re-solving the GateUp family rather than copying the old assignment lowers latency by 11\% and 6\% at M128 and M512, selects the same instance at M2048 and loses 2.3\% at M4096. At attention length 6144, the expression-selected split loses 4.5--16.0\% to a diagnostic split; the table reports the expression-selected program.

\subsection{Knowledge-assisted construction}
We compare construction with the common specification and tools against construction supplied with an executable family, its cost expression and explanatory optimization knowledge. Each condition has three independent agent runs using GPT-6 Astra at medium reasoning effort. Agents can revise the program, decision variables and expression using their own measurement feedback; the final instance is selected by the expression rather than by taking the fastest measured probe. The tasks are batch-one, single-query grouped-query attention at $(N,H,H_{KV},D)=(1025,16,4,64)$, $(6145,32,8,128)$ and $(24577,64,8,128)$, where $N$ is KV length, $H$ and $H_{KV}$ are query and KV head counts, and $D$ is head dimension. Each task uses fresh trajectories; later tasks receive no candidates or feedback from earlier tasks. Timing covers attention and split reduction, including input-dependent transformations, but excludes projection, rotary embedding and KV-cache updates.

Each run has cumulative limits of 2400 seconds for agent construction, 2400 seconds for compilation and measurement, and 24 distinct evaluated source--assignment pairs. For the first task, all runs receive the same continuation opportunity, up to two additional feedback batches within those limits. The second and third tasks receive four feedback batches from the start. Independent agent concurrency is three for the first task and six for the later tasks; GPU measurements remain serialized. The final candidates, a copied development configuration, PyTorch eager and both FlashInfer modes (tensor-core and SIMT) are measured together on GB10 using warmed CUDA Graph replay. All pass the same direct and Graph checks on random, zero, scaled and tail-dominant inputs against an independent FP32 reference.

\begin{table}[!htb]\centering\small
\caption{Knowledge-assisted construction on three attention tasks. Latency is the median [minimum, maximum] across three independent agent runs. Speedup compares that median with the faster FlashInfer mode in the common batch. Recorded active cost is the median sum of agent construction, solving and worker execution time per run.}\label{tab:template-construction}
\begin{tabular}{@{}llrrr@{}}\toprule
Task & Condition & Latency (\us) & vs. FlashInfer & Active cost (s)\\\midrule
$N=1025$ & Task only &5.598 [5.326, 6.400]&0.835$\times$&2225\\
& Executable template &3.142 [3.137, 3.443]&1.488$\times$&2069\\\addlinespace
$N=6145$ & Task only &71.725 [58.424, 72.786]&0.942$\times$&2089\\
& Executable template &58.362 [56.626, 61.149]&1.158$\times$&2083\\\addlinespace
$N=24577$ & Task only &422.976 [419.722, 423.445]&1.038$\times$&1826\\
& Executable template &425.649 [424.966, 468.482]&1.031$\times$&2075\\\bottomrule
\end{tabular}
\tnote{Same-batch eager/FlashInfer/copied references are 28.196/4.674/4.048~\us{} for $N=1025$, 1143.042/67.586/62.740~\us{} for $N=6145$, and 9820.396/439.035/429.582~\us{} for $N=24577$. Costs exclude shared development (1205~s recorded), infrastructure repair and unmeasured historical knowledge and manual engineering. On the first task, earlier feedback for two runs per condition came from a previous host before all runs continued on GB10; final timing uses GB10 only. The cost comparison covers construction of each new family, excluding those shared costs.}
\end{table}

\section{Wall-relaxation protocol}\label{app:wall-details}
\lead{Pricing register capacity recovers the joint solution across objectives.}
We isolate capacity-driven software reselection on Qwen2.5-0.5B at 128 and 512 tokens, using prefill, decode and mixed objectives. Nine prefill and seven decode programs vary tiling, packing, fusion and execution order. The hardware domain contains twelve combinations of vector-register capacity $\{512,1024,2048\}$ and scalar-register capacity $\{128,256,512,2048\}$ slots per SM, under the same 1.11~mm$^2$ area ceiling. Other fields are fixed: two SMs, eight 32-thread warps per SM, 16~KiB shared memory and 32~KiB L1 per SM, 128~KiB L2, and 64~bytes/cycle external-memory service. A program is admitted only if every compiled kernel fits eight resident one-warp blocks. Within this condition, capacity changes enable different programs while leaving the cost of a fixed program unchanged; lower-occupancy mappings are outside this experiment.

The objective is the maximum scenario-specific cycle ratio to the reference program and chip: two lengths for prefill or decode, and all four scenarios for mixed. Fresh instruction simulations provide 36 coefficient measurements, including reference costs of 814.12/3767.13 million cycles for prefill and 155.35/181.55 million for decode. Each result aggregates a representative layer across 24 layers plus the language-model head, with the same decode warm-up convention across methods. Fixed hardware optimizes software at the starting chip; sequential search then changes hardware with that software fixed. Random and joint search consider the twelve chips. Wall relaxation prices capacity increases by predicted objective reduction per added area and reselects software after each purchase. The coupled extension additionally screens pairs of capacity constraints together. Inner problems are exact minima over the finite program menu.

Instruction-simulator evaluation confirms all 57 selected results and ten logical purchases across the two wall variants and three objectives. Sixteen distinct physical endpoints cover the selections and purchase transitions, with identical configurations reused across comparisons. The model uses instruction-simulator coefficients; replaying the purchased configurations reproduces every predicted gain exactly, satisfying the prespecified 5\% tolerance and confirming the capacity-driven software changes. Validation follows model search, so the experiment does not measure an interleaved online optimization loop. Data collection and validation consume 7316 seconds of summed stage execution time, including failed and partial attempts; 2720 seconds of queue waits are recorded separately. Parallel stages may overlap, so their sum is not elapsed time to completion. Individual finite-menu searches take 0.12--2.12~ms after coefficients are available, excluding offline simulation and unmeasured historical development.

\section{Verification across program revisions}\label{app:revision-details}

We next evaluate reuse when optimization changes the compiled program. Full checking reconstructs and solves every obligation. Node reuse retains certificates for unchanged nodes of the dependency graph. Query reuse additionally retains prior UNSAT results when the parsed formula and its checker, specification and solver context match. Changed programs are decoded and their obligations rebuilt before reuse; unresolved results never become proofs.

A controlled sequence contains 16 revisions of cost expressions, instance parameters, program structure and verification premises, including three injected ELF faults, and is replayed at one, two, four and eight shapes. A second sequence contains seven actual agent proposals for an operand-load-batching family. Its complete 59-version history also includes 52 host diagnostic executions. All strategies start with empty reuse state and include initial checking, identity checks, evidence authentication, dependency maintenance and persistent writes. The controlled and seven-proposal sequences have three paired replays; the complete history has one. These are verification replays of fixed program histories, with generation and compilation outside the timing.

\begin{table}[!htb]\centering\small
\caption{Verification cost across program revisions. Times are sequence totals in seconds (medians for three replays); speedups are medians of paired full/reuse ratios, so they need not equal ratios of displayed medians. Query reuse includes node reuse. A dash denotes an unmeasured strategy. The final row is one replay; all other rows have three.}\label{tab:verification-revisions}
\setlength{\tabcolsep}{4pt}
\begin{tabular}{@{}lrrrrrr@{}}\toprule
Sequence & Versions & Full (s) & Node (s) & Query (s) & Full/Node & Full/Query\\\midrule
Controlled, 1 shape & 16 & 4.06 & 2.34 & -- & 1.72$\times$ & --\\
Controlled, 2 shapes & 16 & 6.48 & 3.94 & -- & 1.66$\times$ & --\\
Controlled, 4 shapes & 16 & 11.74 & 7.24 & -- & 1.62$\times$ & --\\
Controlled, 8 shapes & 16 & 25.19 & 15.87 & -- & 1.59$\times$ & --\\\midrule
agent proposals & 7 & 10.74 & 10.56 & 8.56 & 1.02$\times$ & 1.25$\times$\\
Complete execution history & 59 & 83.04 & 81.66 & 72.17 & 1.02$\times$ & 1.15$\times$\\\bottomrule
\end{tabular}
\end{table}

All strategies preserve the same classifications on the real histories; full and node reuse also agree on all 192 controlled comparisons. A separate 16-revision fault replay preserves the expected outcomes. Of the 59 real versions, 45 are proved within the fixed-shape, mathematical-real checker for compiled Executable and Linkable Format (ELF) programs and 14 remain unresolved due to unsupported spills of thread-private values to memory. This extends the reuse evidence from adding refinement levels to modifying implementations, while keeping proof reuse distinct from floating-point numerical validation. The trajectory also informed development of query reuse; independent optimization histories remain a further transfer test.

\section{Cost-model fidelity and chip transfer}\label{app:model-fidelity}

A Qwen2.5-0.5B study searches two software families and 27 hardware fields under a 1.11~mm$^2$ area ceiling, using worst-scenario cycle ratios over prefill lengths 128/512 and 16 decode steps from histories 128/512. Nine hardware-only, sequential and joint optimizations reach zero solver gap and predict ratios near 0.333; complete instruction simulation gives 0.791--0.899. On prefill, the joint choice is slower than hardware-only (0.899 versus 0.791), exposing a ranking error. In 18 transfer evaluations, Qwen-selected chips take 3.07--5.60$\times$ the baseline cycles of GPT-2 small and Pythia-160M.

\section{Structural and workload checks}\label{app:refinement-checks}

The Python/Z3 checker evaluates 78 variants across eight structural groups, from reduction partitioning and buffer reuse to resource scheduling and two connected GEMMs. It identifies 31 new faults and three deadlocks through counterexamples replayed from initialization, while distinguishing inherited faults, failed certificates and infeasible domains. Workload checks accept 36 causal-attention schedules, 48 prefill-and-decode schedules and 16 training-transition schedules, and replay counterexamples for all 3/6/7 injected fault families, respectively.

\FloatBarrier
\endgroup

\end{document}